\def\year{2022}\relax
\documentclass[letterpaper]{article} 
\usepackage{aaai22}  
\usepackage{times}  
\usepackage{helvet}  
\usepackage{courier}  
\usepackage[hyphens]{url}  
\usepackage{graphicx} 
\usepackage{natbib}  
\usepackage{caption} 
\usepackage{helvet}
\usepackage{courier}
\usepackage{subfigure}
\usepackage{graphicx}
\usepackage{multirow}
\usepackage{booktabs}
\usepackage{tabularx}
\usepackage{amsthm,amsmath,amssymb}
\usepackage{mathrsfs}
\usepackage{bm}
\usepackage{verbatim}
\DeclareCaptionStyle{ruled}{labelfont=normalfont,labelsep=colon,strut=off} 
\usepackage{algorithm}
\usepackage{algorithmic}
\DeclareMathOperator*{\argmax}{arg\,max}

\usepackage{newfloat}
\usepackage{listings}
\floatstyle{ruled}
\newfloat{listing}{tb}{lst}{}
\floatname{listing}{Listing}

\title{Feudal Multi-Agent Reinforcement Learning with Adaptive Network Partition for Traffic Signal Control}
\author {
    Jinming Ma,\textsuperscript{\rm 1}
    Feng Wu, \textsuperscript{\rm 1}
}
\affiliations {
    \textsuperscript{\rm 1} School of Computer Science and Technology, University of Science and Technology of China, CHN\\

    jinming@mail.ustc.edu.cn, wufeng02@ustc.edu.cn
}

\begin{document}
	
	\maketitle
	
	\begin{abstract}
		Multi-agent reinforcement learning (MARL) has been applied and shown great potential in multi-intersections traffic signal control, where multiple agents, one for each intersection, must cooperate together to optimize traffic flow. To encourage global cooperation, previous work partitions the traffic network into several regions and learns policies for agents in a feudal structure. However, static network partition fails to adapt to dynamic traffic flow, which will changes frequently over time. To address this, we propose a novel feudal MARL approach with adaptive network partition. Specifically, we first partition the network into several regions according to the traffic flow. To do this, we propose two approaches: one is directly to use graph neural network (GNN) to generate the network partition, and the other is to use Monte-Carlo tree search (MCTS) to find the best partition with criteria computed by GNN. Then, we design a variant of Qmix using GNN to handle various dimensions of input, given by the dynamic network partition. Finally, we use a feudal hierarchy to manage agents in each partition and promote global cooperation. By doing so, agents are able to adapt to the traffic flow as required in practice. We empirically evaluate our method both in a synthetic traffic grid and real-world traffic networks of three cities, widely used in the literature. Our experimental results confirm that our method can achieve better performance, in terms of average travel time and queue length, than several leading methods for traffic signal control. 
	\end{abstract}
	
	\section{Introduction}
	Nowadays, the growing traffic congestion has brought serious negative impacts on environmental protection, urban economic development and our daily lives. To meet the increasing traffic demand, many efforts have been made to optimize traffic control and improve road capacity. Traditional traffic signal control (TSC) mostly relies on fixed timing strategy, which cycles the established settings periodically \cite{koonce2008traffic}. However, those methods, which heavily depend on expert knowledge and heuristic assumptions, have many weak points. For example, they often cause long traffic delays and are unable to make flexible adjustments based on real-time traffic information. In recent years, researchers \cite{van2016coordinated, wei2019survey, chu2019multi} have tried to use deep reinforcement learning (DRL) for TSC and shown better performance than traditional TSC methods.
	
	To date, there are several DRL based methods \cite{prashanth2010reinforcement, wei2018intellilight} that control each intersection independently and have no coordination with other intersections. Indeed, the lack of coordination among intersections will lead to poor efficiency of the overall traffic flow. With regards to this, many studies \cite{wiering2000multi, kuyer2008multiagent} propose to solve TSC using multi-agent RL (MARL), which trains agents to cooperate in a centralized or decentralized manner. Generally, decentralized methods have better scalability than centralized approaches because each agent independently learns its own policy. However, due to partial observability of the agents, decentralized MARL may get stuck in local optima more easily.
	
	To address this, researchers have investigated several techniques to approach global optima. The most common way is to share observation and fingerprint between adjacent agents for stable cooperative control \cite{chu2019multi}. Some work uses graph attention networks to facilitate additional information from neighboring intersections to optimize the traffic \cite{wei2019colight}. Others \cite{wei2019presslight,chen2020toward} are based on max-pressure theory to implicitly encourage the cooperation between neighboring intersections. Most recently, FMA2C \cite{MWaamas20} was proposed to extend MA2C \cite{chu2019multi} with a feudal structure and improve global coordination among agents. Specifically, it first splits the traffic network into several regions. Then, a manager-worker hierarchy is constructed, where each region is assigned to a manager and the intersections in the region are controlled by its workers. In this way, managers can cooperate more globally at the regional level and guide the coordination of their workers. However, a limitation of this approach is that the regions are split manually and fixed afterward. This becomes inefficient when the pattern of traffic flow changes frequently.
	
	In this paper, we propose a novel feudal MARL with adaptive network partition for multi-intersection TSC. Specifically, given the underlying traffic network, we first construct the dynamic flow network to model the traffic flow at different periods. Then, we introduce two network partition methods based on graphical neural network (GNN) and Monte-Carlo tree search (MCTS) respectively. For the former, we directly use GNN to generate the network partition according to the traffic flow. This approach is easy to implement and can be trained in an end-to-end manner. However, it is hard to train and may generate unreasonable partition for a large network. In MCTS, we find the best partition through an explicit search process and can scale to a large network. To facilitate different network partitions, we design a variant of Qmix \cite{rashid2018qmix} to handle various dimensions of input. Similar to FMA2C, we use the feudal hierarchy where manager agents cooperate at the high-level and communicate its sub-goals to lower-level worker agents in the region. In the experiments, we tested our algorithm both in a synthetic traffic grid and real-world traffic networks of three cities. Our experimental results demonstrate that the proposed method benefits from the adaptive network partition and significantly outperforms several state-of-the-art methods for TSC.
	
	\section{Background}
	In this paper, we consider the TSC problem in a multi-intersection network as illustrated in Figure 1(a). Here, each intersection has 12 traffic movements, consisting of four approaches and three lanes (turning right, turning left and going straight) per approach. As shown in Figure 1(b), the green dot indicates that the movement is allowed and several non-conflicting movements can be combined in a phase. The objective of TSC is to choose the optimal phase of each intersection to maximize the traffic flow. As in the literature, this problem can be modeled as a Markov game, where each intersection in the traffic network is controlled by an agent.
	
	\subsection{Markov Game for Traffic Signal Control}
	Specifically, the Markov game is defined by a tuple $\left\langle \mathcal{S}, \mathcal{O}, \mathcal{A}, \mathcal{T}, \mathcal{R} \right\rangle$. $\mathcal{S}$ is a finite set of states, and each agent can observe a part of the state $s \in \mathcal{S}$ as its observation $o \in \mathcal{O}$. $\mathcal{A}$ is a set of actions for each agent. $\mathcal{T}$ is the transition function with joint action of all agents $\textbf{a}$. Each agent obtains an immediate reward $r$ by a reward function $\mathcal{R}(s, \textbf{a})$.
	
	For TSC, each agent observes the quantitative descriptions (i.e. observation) of its intersection, such as queue length, waiting time and delay. Then, the agent chooses a phase (i.e. action) for the intersection and receives an immediate reward $r$ from the environment, which indicates the traffic situation. The goal is to optimize the situation of the overall traffic network. This can be done by MARL. Comparing with the conventional TSC methods that heavily rely on pre-defined rules, the main advantage of solving TSC by MARL is that agents can learn a better policy by directly interacting with the real-time and dynamic environment.
	
	\begin{figure}[t]
		\centering
		\subfigure[Multi-intersection traffic network in Hangzhou, China.] {\includegraphics[width=0.45 \hsize]{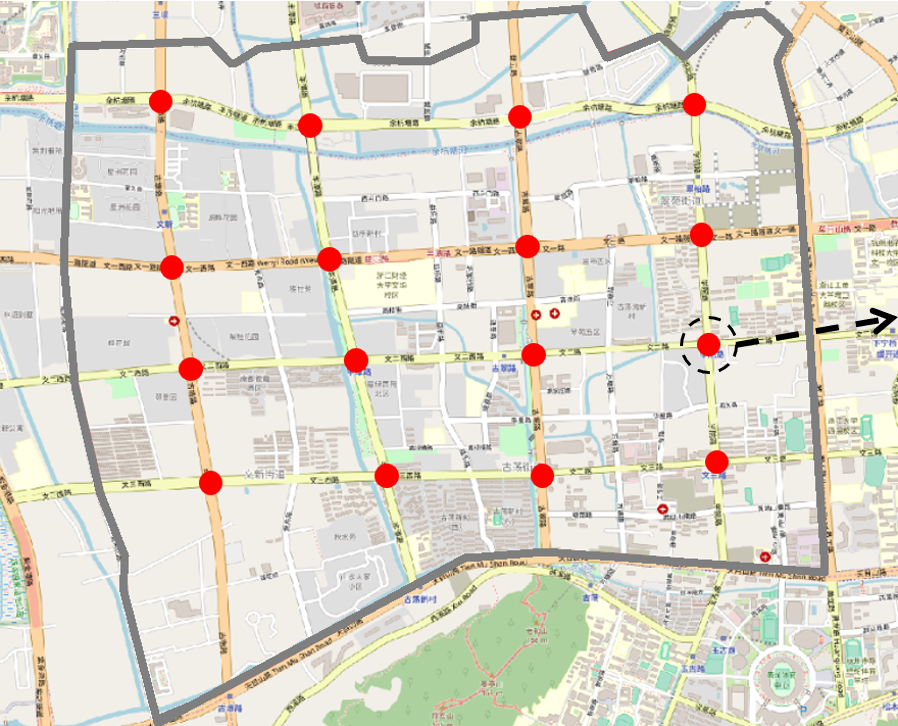}}
		\subfigure[Intersection and its 12 movements.] {\includegraphics[width=0.4 \columnwidth]{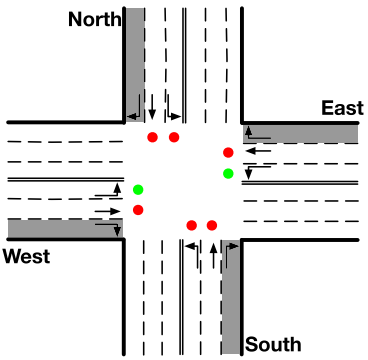}}
		\caption{Example of the traffic network and intersection.}
		\label{network}
	\end{figure}
	
	\subsection{Multi-Agent RL for Traffic Signal Control}
	
	To date, many efforts have been made to design algorithms for cooperating agents in scenarios of TSC. One typical approach is to train a central agent to control all intersections \cite{prashanth2010reinforcement}, or jointly model the action among learning agents with centralized optimization \cite{kuyer2008multiagent, van2016coordinated}. Unfortunately, due to the curse of dimensionality, these {\em centralized} methods usually have the {\em scalability} issue and therefore are hard to apply on large-scale road networks.
	
	Other studies aim to encourage agents to learn cooperation with others in a {\em decentralized} way by adding information from neighbor \cite{aziz2018learning, wei2019colight, chu2019multi} or designing an implicit cooperation mechanism that induces cooperation between agents \cite{wei2019presslight, xu2021hierarchically}. However, due to the problem of partial observability, these methods may have difficulty achieving {\em global optimal control}.
	
	To alleviate this, a recent approach models the problem as a feudal Markov game to improve global coordination among agents with a worker-manager hierarchy \cite{MWaamas20}. For worker agent, it controls an intersection by select a phase as in standard multi-agent RL. For manager agents, it controls a region, which includes agents of multiple intersections, by setting sub-goals for them. Intuitively, manager agents with a higher-level perspective of the network can make more global decisions.
	
	Although the feudal hierarchy does improve global coordination among multi-intersections, the network partition (i.e., which region is controlled by a manager) is {\em static} and not flexible when the traffic pattern changes frequently. As we observed in the experiments, when traffic patterns change, the hierarchy becomes less effective if the congested area spans two static regions. To address this, we propose a novel RL approach with {\em adaptive} network partition that is more flexible and can adapt to complex traffic patterns.
	
	\section{Method}
	
	\begin{figure*}[t]
		\centering
		\includegraphics[width=1.9\columnwidth]{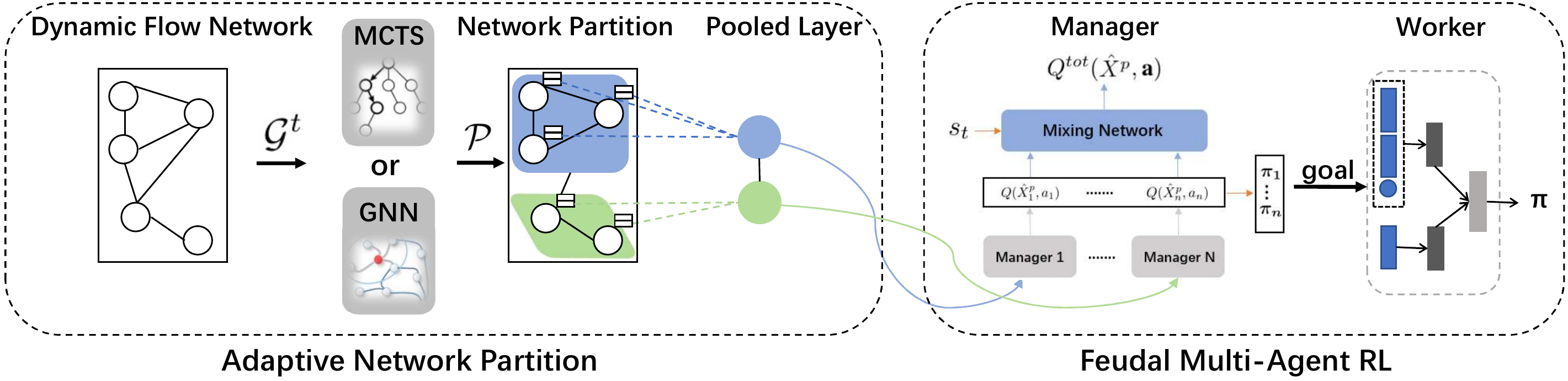}
		\caption{The overall framework of AFMRL.}
		\label{network2}
	\end{figure*}
	
	In this section, we propose AFMRL | Adaptive Feudal Multi-agent Reinforcement Learning with dynamic network partition for TSC. The overall framework of our method is shown in Figure \ref{network2}. From left to right, we start with a formal definition of dynamic flow network and introduce two adaptive network partition methods based on GNN and MCTS respectively. For the former, we directly use GNN to generate the network partition. For the latter, we use MCTS to search the best network partition whose value is measured by GNN. Then, we designed a variant of Qmix to process different dimensional inputs given the network partition. After that, we construct a feudal hierarchy to learn policies for the regions and finally for each intersection.
	
	\subsection{Adaptive Network Partition}
	We model a traffic network as directed graph $G(\mathbb{V}, \mathbb{E})$, where each vertex $v \in \mathbb{V}$ represents an intersection and each edge $e_{ij} \in \mathbb{E}$ represents the road from intersection $i$ to $j$. Now, we introduce the dynamic flow network to capture the dynamic traffic flow that changes over time. Specifically, the {\em dynamic flow network} is defined as  $\mathcal{G}^t(G, F^t)$, where: $G$ is the traffic network; $f_{ij}^t = (e_{ij}, w) \in F^t$ is the traffic flow in edge $e_{ij}$ with traffic density descriptions $w$, which is ratio of queue length over the capacity of the road.

    Given the dynamic flow network, we partition the network $\mathcal{G}^t$ into $m$ disjoint regions $\mathcal{P}$ = $\{P_1, ..., P_m\}$, where ${\forall}_{P_i,P_j}, \cup^m_{k=1} P_k = \mathcal{G}, P_i \cap P_j = \emptyset$. Here, at different timestep $t$, the dynamic flow network can be partitioned into different disjoint regions depending on the real-time traffic flow. As aforementioned, our goal is to adaptively partition the network so that it is beneficial for the learning algorithm.
    
    To this end, we propose two approaches: one is to directly generate the network partition through GNN, and the other is using MCTS to search the network partition with criteria computed by a separate GNN. The advantage of the former is that it can train the entire network in an end-to-end fashion. However, it may fail to generate a reasonable partition for larger network. On the contrary, the latter can generate a good network partition through explicit search in a feasible time. We will describe both methods in the following sections and compare their performance in the experiments.
	
	\subsubsection{GNN for network partition.} 
	
	

    Here, we introduce our approach by applying the GNN to output the network partition $\mathcal{P}$ for the dynamic flow network. Inspired by \citet{ying2018hierarchical}, we compute the assignment matrix $\mathcal{M}$ of agents using a \emph{pooling} GNN that takes the input observation of intersections $\mathcal{O}$ and the real-time flow $F^t \in \mathbb{R}^{n_a \times n_a}$ as follow:
    \begin{equation}
        \label{assignment}
        \mathcal{M} = softmax(GNN_{pool}(F, \mathcal{O})) \in \mathbb{R}^{n_a \times n_p}
    \end{equation}
    where $n_a$ is the number of intersections,  $n_p$ is the number of regions. Given this, we can get the network partition $\mathcal{P}$ through the assignment matrix $\mathcal{M}$, i.e., region $P_i$ consists of the agents whose $i$-th column of the matrix is maximum.

    In practice, it is generally difficult to train the GNN using only gradient signal from the RL in an end-to-end fashion as shown in Figure \ref{network2}. Hence, we use an auxiliary link loss to encode nearby nodes that should be pooled together as suggested by \citet{ying2018hierarchical}. Even with this improvement, we observed that GNN may not provide the best network partition, because the complex topological structure of graphs is hard to learn with limited feedback. It becomes more severe on a larger traffic network, which will be shown in our experiments. This motivates us to propose an alternative approach based on MCTS as described next.
	
	\subsubsection{MCTS for network partition.} 
	
\begin{figure}[t]
		\centering
		\includegraphics[width=0.9 \hsize]{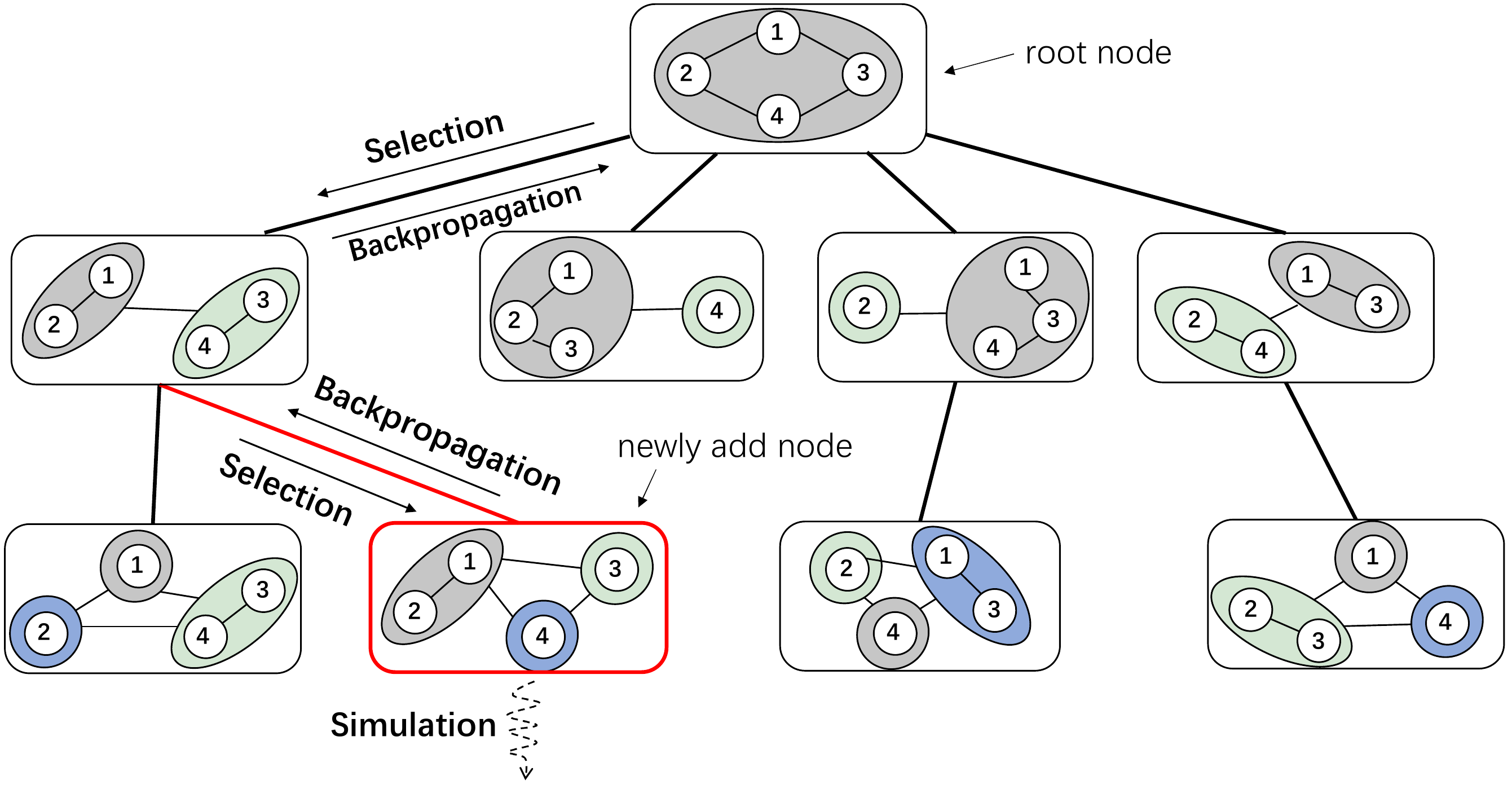}
		\caption{Illustration of partial search tree built by MCTS for network partition.}
		\label{st}
	\end{figure}

    Now, we turn to propose our MCTS method to search the best network partition in the dynamic flow network. Inspired by \citet{WRijcai20}, we model the partition problem as a search process over tree nodes. As shown in Figure \ref{st}, each tree node is associated with a network partition and the root node of the search tree is the original network containing all agents, e.g., $\mathcal{P}$ = $\{\{1,...,n\}\}$. The children of each node are obtained by bipartition of one region of its parent (e.g., a child of the root node is $\{\{1,...\}, \{...,n\}\}$). Each branch of the tree expands until the termination condition meets at a leaf node. When all branch reaches its leaf, the search is completed.
    
    We incrementally build the search tree iteration by iteration. In each iteration, as shown in Figure \ref{st}, we start the search from the entire network down to split network and expand the search tree based on four steps of MCTS. In the {\em selection} step, an optimal child node is selected among all children so that the tree can expands to the subspace where the optimal solution is most promising. Similar to previous MCTS methods, we use the UCB1 heuristic \cite{auer2002finite} to select branches as:
    \begin{equation}
        \label{UCB }
        UCB1(\mathcal{P}') = V'(\mathcal{P}') + c \sqrt{\frac{log N(\mathcal{P})}{N(\mathcal{P}')}}
    \end{equation}
    where $V'(\mathcal{P}) = max_{\mathcal{P}' \in Tree'(\mathcal{P})} V(\mathcal{P}')$ is the current maximal value with the current search space $Tree'(\mathcal{P})$, N($\mathcal{P}$) is the frequency that the node $\mathcal{P}$ is visited when searching, and $c$ is a constant parameter. Given this, we can select the child node that maximizes the UCB1($\mathcal{P}'$), i.e., $\mathcal{P}^* = \argmax_{\mathcal{P}' \in Child(\mathcal{P})} UCB1(\mathcal{P}')$. In the {\em expansion} step, if the selected child $\mathcal{P}^* \in Child(\mathcal{P})$ is currently not in the tree, we expand by adding a new node $\mathcal{P}^*$ as a child of $\mathcal{P}$.

    In the {\em simulation} step, to evaluate the value $V'(\mathcal{P}^*$) of the new node $\mathcal{P}^*$, we use the default policy to perform a rollout search by successively bipartitioning a region until the termination condition is meet. The value of $V'(\mathcal{P}^*$) is initialized by $V'(\mathcal{P}^*) = max_{\mathcal{P}' \in Trace(\mathcal{P}^*)} UCB1(\mathcal{P}')$, where $Trace(\mathcal{P}^*)$ is a set of partitions encountered during the rollout search from $\mathcal{P}^*$. In rollout search, for $\mathcal{P}$, the default policy to perform network partition is to select the partition $\mathcal{P}' \in Child(\mathcal{P})$ with maximizing the value $V(\mathcal{P}')$. In the {\em backpropagation} step, after the simulation of $\mathcal{P}^*$, its all ancestors will updated by backpropagting the value $V'(\mathcal{P}^*$). For each ancestor node, its value is updated by $V'(\mathcal{P}^*$), which is the optimal solution in the sub-tree rooted by it.
	
	Each iteration expands the search tree, and as the number of iterations increases, the size of the search tree continues to increase. When all nodes are added to the tree or time runs out, the complete solution space has been searched. In the former case, we can search the optimal network partition while in the later case the currently best solution is searched.
	
	
	\subsubsection{Network partition value.} 
	
	
    In MCTS, one remaining challenge is to define the value function $\mathcal{V}(\mathcal{P})$ of $\mathcal{P}$ for any network partition $\mathcal{P}$. With this, the goal of MCTS is to find the most valuable partition $\mathcal{P}^*$ in the set of feasible partition denoted by \textbf{$\mathbb{P}$}, i.e., $\mathcal{P}^*$ = $\argmax_{\mathcal{P} \in \mathbb{P}} \mathcal{V}(\mathcal{P})$. Unfortunately, this value function is unknown in our setting and must be learned along with the RL algorithm. Here, we use GNN to extract useful feature embeddings from the input graphs, including vertex, edge and topology information, and predict the network partition value. Note that this is a separate GNN that is structurally different from the 
   aforementioned GNN for group partition.
	
	\begin{figure}[t]
		\centering
		\includegraphics[width=0.9 \hsize]{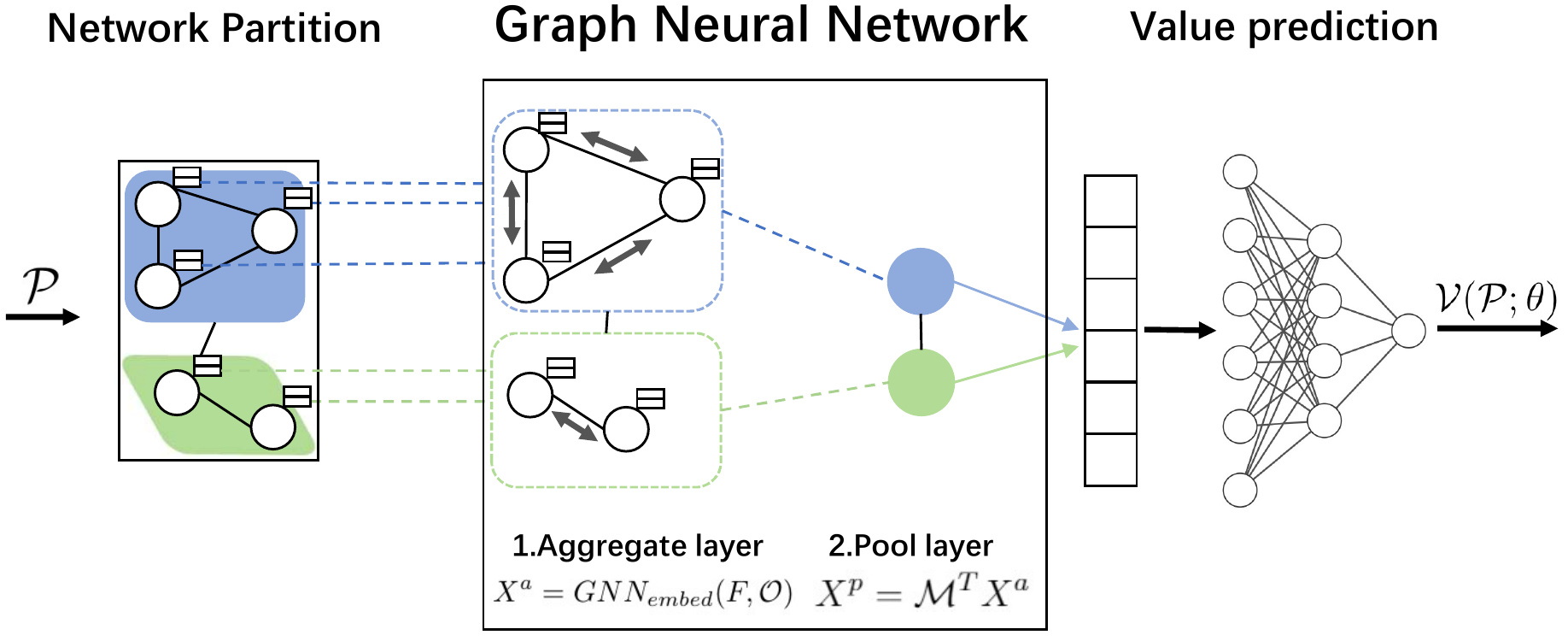}
		\caption{Illustration of predicting the network partition value $\mathcal{V}(\mathcal{P};\theta)$ by GNN. The colored rectangle represents a region, and the black line represents the traffic flow between them. The colored dotted line represents the process according to Equations \ref{embedding1} and \ref{embedding2}.}
		\label{network1}
	\end{figure}

    As shown in Figure \ref{network1}, the network takes the observation of agents $\mathcal{O}$, the real-time flow $F^t$ and the assignment matrix $\mathcal{M} \in \mathbb{R}^{n_a \times n_p}$ as input. We first generate the embeddings $X^a$ of agents by an \emph{embedding} GNN module \cite{ying2018hierarchical}, which is applied to $\mathcal{O}$ and $F$ as below:
    \begin{equation}
        \label{embedding}
        X^a = GNN_{embed}(F, \mathcal{O}) \in \mathbb{R}^{n_a \times d}
    \end{equation}
    where $d$ is the dimension of agents' embeddings. In this step, the GNN module performs a relation reasoning that propagates information across edges of the graph. Then, we generate a coarsened adjacency matrix $\mathcal{F}$ denoting the connectivity strength between each pair of regions (i.e., the traffic flow between regions), and the embeddings $X^p$ for regions by pooling these agents' embeddings in the region together according to the assignment matrix $\mathcal{M}$:
    \begin{equation}
        \label{embedding1}
        \mathcal{F} = \mathcal{M}^T  F \mathcal{M}  \in \mathbb{R}^{n_p \times n_p}
    \end{equation}
    \begin{equation}
        \label{embedding2}
        X^p = \mathcal{M}^T  X^a \in \mathbb{R}^{n_p \times d}
    \end{equation}
    Similarly, we take the relational matrix between the regions $\mathcal{F}$ and their embeddings $X^p$ through an \emph{embedding} GNN module to get new embeddings ${\hat{X}^p}$ for the regions:
    \begin{equation}
        \label{embedding3}
        {\hat{X}^p} = GNN_{embed}(\mathcal{F}, X^p) \in \mathbb{R}^{n_a \times d}
    \end{equation}
    Finally, the new embeddings feed in the neural network to output the value $\mathcal{V}(\mathcal{P};\theta)$. The network is trained in an end-to-end fashion by minimizing $\mathcal{L}$ = MSE($\mathcal{V}(\mathcal{P};\theta), r$), where $r$ is the reward received from environment by forming the network partition $\mathcal{P}$.

	
	
	Since the GNN requires a learning process, its prediction value may not be useful for MCTS especially in the early stage. Note that the learning process of GNN and the search process of MCTS are inter-dependent: GNN needs good partitions to learn their values while MCTS requires accurate values to find the good partition. Therefore, we introduce a heuristic function to boost the learning process of GNN.
	
	
	We borrow ideas from graph theory \cite{chu2016large} to evaluate how strong the connection is between two regions: $cut(P_i, P_j)=\sum_{u \in P_i, v \in P_j} f(u, v)$ for any $\{P_i, P_j\}$, where $f(u, v)$ is the traffic flow between intersections $u$ and $v$. Intuitively, a good partition should minimize this value between every pair of regions. Furthermore, to avoid undesirable bias for partitioning out small sets, we use ``Normalized cut'' \cite{shi2000normalized} to measure the disassociation of the network partition:
	
	\begin{equation}
		\label{Ncut}
		Ncut(P_1, ..., P_m) = \sum_{i, j=1 , i \neq j}^{m} \frac{cut(P_i, P_j)} {assoc(P_i)}
	\end{equation}
	where $assoc(P_i)=\sum_{u \in P_i, v \in \mathbb{V}} f(u, v)$ is the total connection from nodes in $P^i$ to all nodes in $\mathcal{G}$. By minimizing $Ncut$, we will find the apposite network partition where the connection within each region is high while the connection between each region is low. Intuitively, this meets our goal of partitioning the traffic networks.

    Given this, we combine the heuristic with the prediction value and compute the value of a network partition $V(\mathcal{P})$ as:
    \begin{equation}
        \label{Partition value }
        V(\mathcal{P}) = (\alpha - 1) Ncut(\mathcal{P}) + \alpha \mathcal{V}(\mathcal{P})
    \end{equation}
    where $\alpha$ is a adaptive weighting coefficient, which is low in the early stage of training and gradually grows along with the learning process.
	
	
	\subsection{Feudal Multi-Agent Reinforcement Learning}
	
	
	Here, we follow the FMA2C framework \cite{MWaamas20} and form a feudal hierarchy with managers and workers, where a manager controls a region in the partition and the agents in the region are its workers who directly controls the traffic intersection. In more detail, the manager is tasked with maximizing the long-term, global reward, learns to communicate sub-goals to multiple simultaneously operating workers. The workers need to learn how to act based on the managerial reward according to the completion of the sub-goals, in addition to maximizing immediate rewards that they experience from the environment.
	
	\subsubsection{Qmix for managers.}

    As aforementioned, each manager controls a region and coordinates the workers inside the region. Note that our objective is to optimize the traffic flow of the entire network. Therefore, the managers of different regions also need to cooperate with each other. To achieve this, we employ Qmix \cite{rashid2018qmix} to learn managers' policies. Notice that, for the different situations of the traffic network, there will be different network partitions, i.e., the number of regions and number of agents in the region. Therefore, Qmix must be modified to handle this issue.

    Specifically, we use a GNN to handle this dynamic dimensions of input. Note that this is a separate GNN with the same structure as shown in Figure \ref{network1}. In more details, we compute the feature embeddings ${\hat{X}^p}$ of regions by Equations \ref{embedding}-\ref{embedding3}, where Equation \ref{embedding3} stabilizes the training process by one-step relational reasoning. Note that, for empty regions, their feature embeddings are $\bf{0}$.
	

    Given the GNN, we use the $m$-th row of ${\hat{X}^p}$ as the feature of region $m$ and feed it into a value network that outputs the individual value function $Q^m$. Then, we put those individual value functions into the mixing network to produce the values of $Q^{tot}$. The hyper-parameter of the mixing network takes the environment state $s$ as input and generates the non-negative weights and bias of the layer in the mixing network. The network is trained in an end-to-end manner using the following loss:
	\begin{equation}
		\label{loss}
		\mathcal{L}(\theta) = \frac{1}{2|B|} \sum_{i}^{|B|} (y_{i}^{tot} - Q_{i}^{tot}({\hat{X}^p}, \textbf{a}, s;\theta))^2
	\end{equation}
	where $B$ is the transitions sampled from the replay buffer, $y^{tot}_t = r_t + \gamma max_{\textbf{a}_{t+1}} Q^{tot}({\hat{X}^p}_{t+1}, \textbf{a}_{t+1}, s_{t+1};\theta^-)$.
	
	\begin{figure*}[t]
		\centering
		\subfigure[Synthetic 4x4] {\includegraphics[width=0.9\columnwidth]{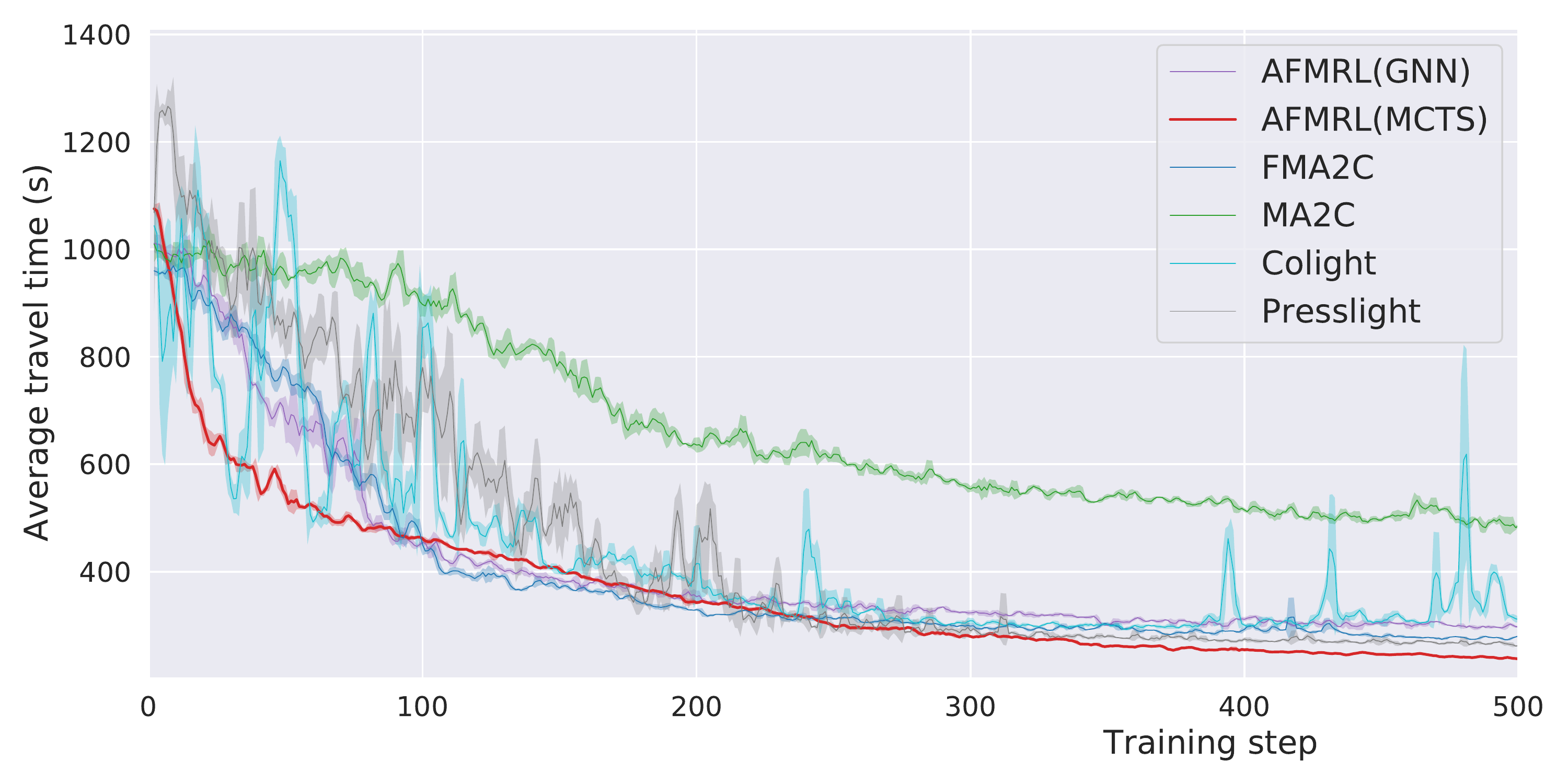}}
		\subfigure[Jinan] {\includegraphics[width=0.9\columnwidth]{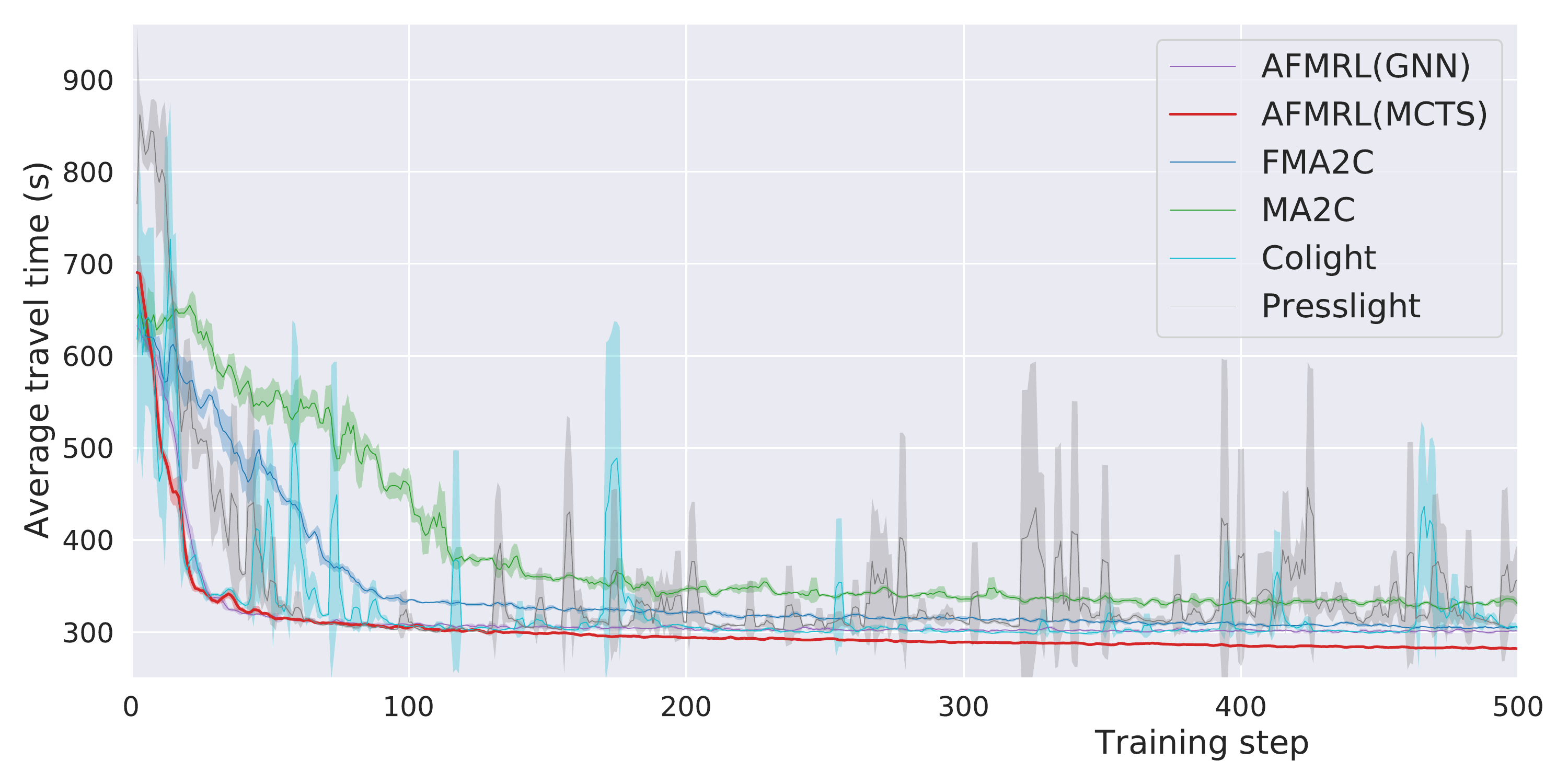}}
		\subfigure[Hangzhou] {\includegraphics[width=0.9\columnwidth]{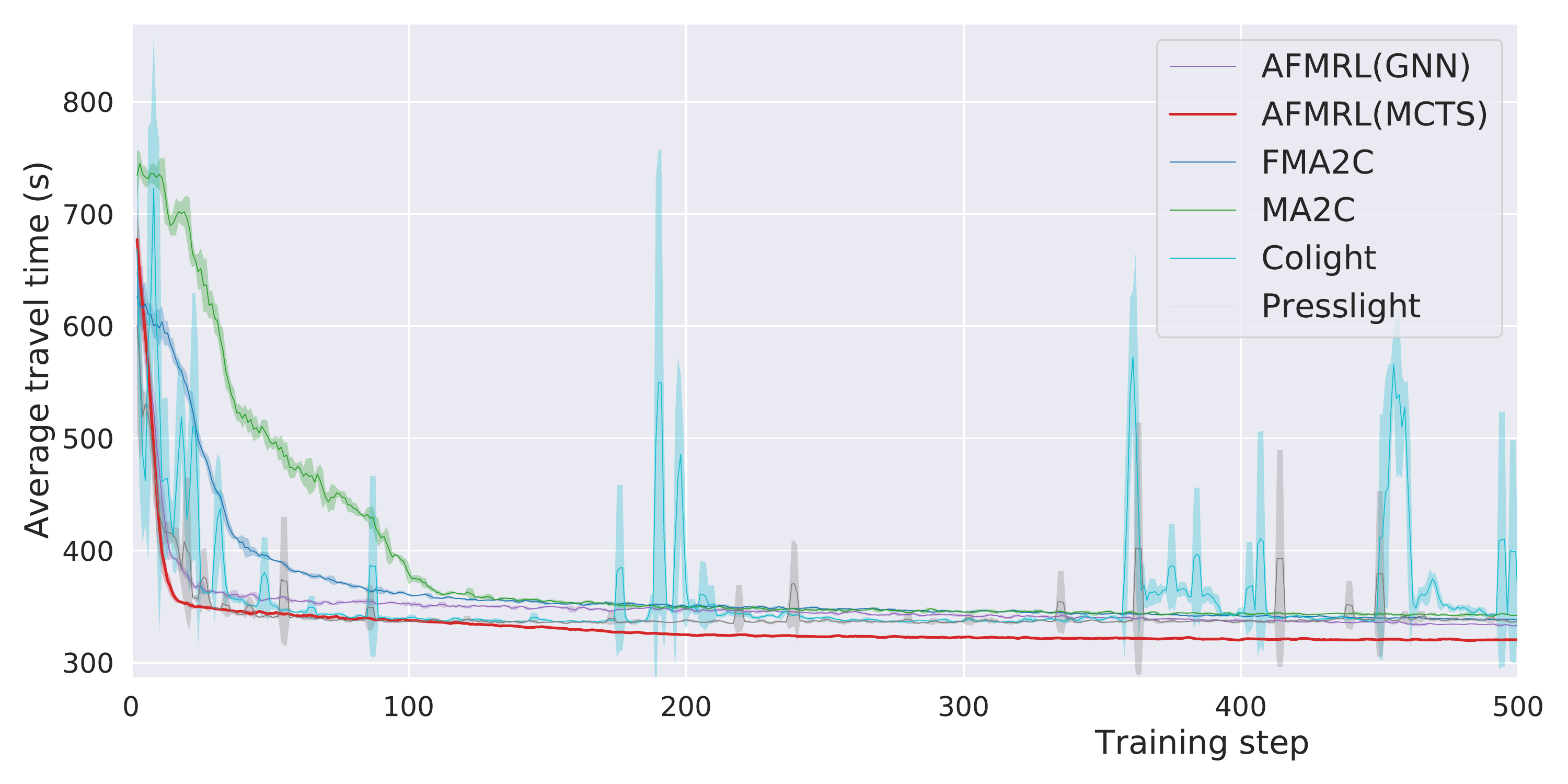}}
		\subfigure[Manhattan] {\includegraphics[width=0.9\columnwidth]{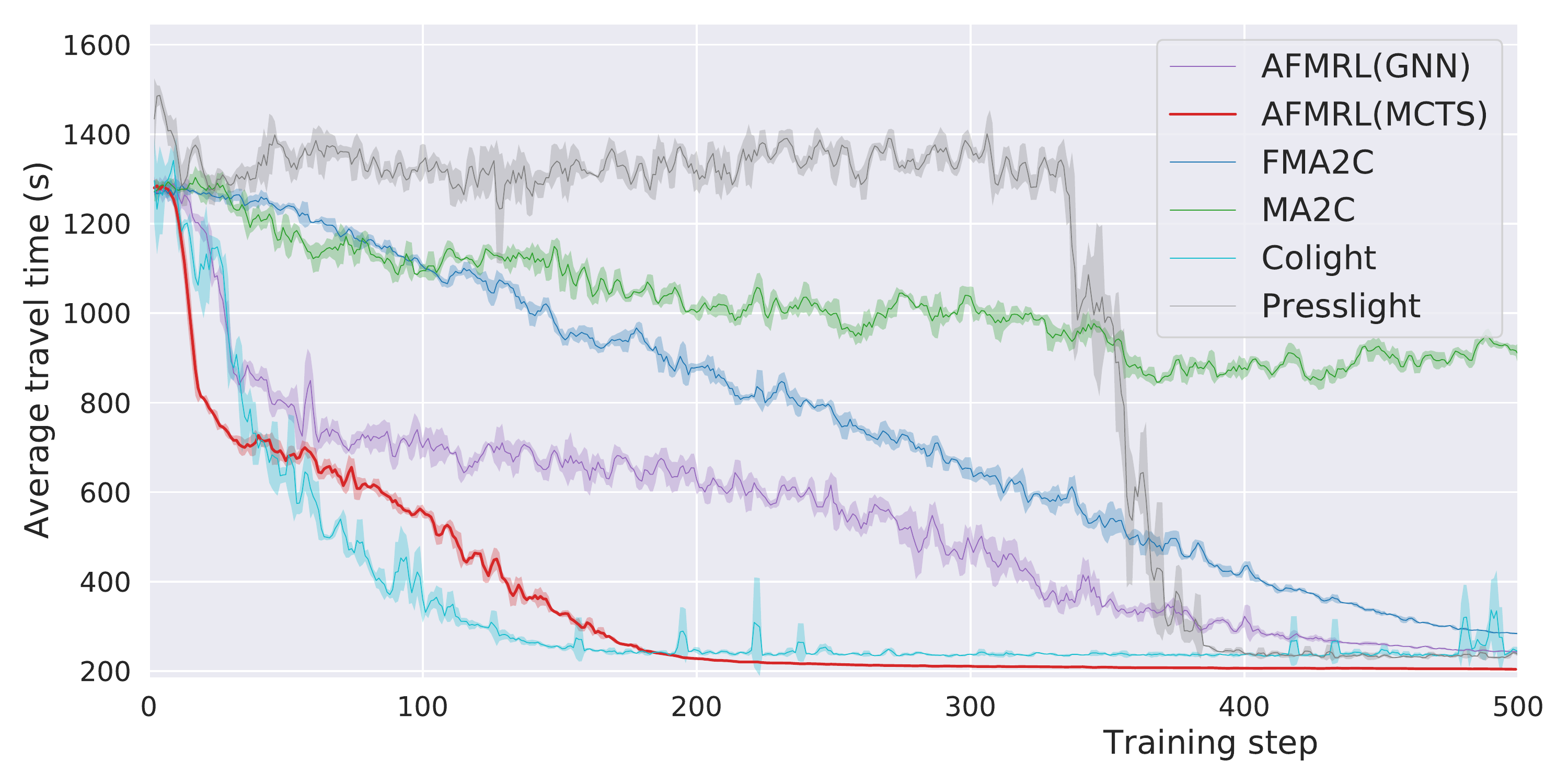}}
		\caption{The train curves of average travel time(s) (The solid line and shade are mean and standard deviation respectively).}
		\label{Training-curves}
	\end{figure*}
	
	
	
	\subsubsection{MA2C for workers.}
	
	Similar to FMA2C \cite{MWaamas20}, we train the policies of workers by the A2C algorithm.  Specifically, at time step $t$, each manager $k$ samples an action $a^M_{t,k}$ from its current policy $\pi^M$ to produce a sub-goal $g$ for the workers in its region. Then, for each worker $i$, we adjust its reward by including an intrinsic reward function $\sigma$ to encourage the worker to complete the sub-goal, such as: $\hat{r}_{t,i} = r_{t,i} + \sigma(o^i_t, g)$, where $r_{t,i}$ is the reward received from the environment and $\sigma$ is the intrinsic reward function mapping from the worker's observation and sub-goal $g$ to a real number. Given the new reward, each worker can learn its policy $\pi^W$ using the A2C algorithm.

	\section{Experiments}
	
	
	\begin{center} 
		\begin{table*}[t]
			\caption{Performance in synthetic 4x4, Jinan, Hangzhou and Manhattan (Best values are in bold).}
			\label{tab:perf}
			\centering\small
			\begin{tabular}{l | c c c c | c c c c}
				\toprule
				\multirow{2}{*}{Method} & \multicolumn{4}{c}{Average Travel Time [s]} & \multicolumn{4}{c}{Avg. queue length [veh]}\\ [5pt]
				\cline{2-9}
				& Synthetic $4\times4$ & Jinan & Hangzhou & Manhattan & Synthetic 4x4 & Jinan & Hangzhou & Manhattan \\ [2pt]
				\midrule
				SOTL & 506.72 & 423.39 & 619.03 & 1745.09 &    76.01 & 26.71 & 23.56 & 88.51\\ [2pt]
				MaxPressure & 251.12 & 337.36 & 374.78 & 305.96 &    27.60 & 13.96 & 6.13 & 8.91\\ [2pt]
				\midrule
				CoLight & 275.55 & 331.59 & 341.61 & 238.35 &    31.87 & 12.08 & 3.13 & 4.19\\ [2pt]
				PressLight & 262.28 & 309.05 & 336.47 & 231.21 &    29.47 & 9.46 & 2.89 & 3.85\\ [2pt]
				MA2C & 421.38 & 328.83 & 340.96 & 351.86 &    53.51 & 12.73 & 3.10 & 5.56\\ [2pt]
				FMA2C & 265.34 & 302.74 & 334.95 & 247.91 &     28.90 & 10.12 & 3.03 & 4.98 \\ [2pt]
				\midrule
				AFMRL(GNN) & 274.37 & 299.25 & 329.52 & 223.10 &     31.73 & 8.77 & 2.68 & 3.52 \\ [2pt]
				\bf{AFMRL(MCTS)} & \bf{ 228.96} & \bf{278.95} & \bf{318.21} & \bf{179.65} &  \bf{23.36} & \bf{5.56} & \bf{1.63} & \bf{1.73}\\ [2pt]
				\bottomrule
			\end{tabular}
		\end{table*}
	\end{center}

    To evaluate the performance of our method, we compared with several conventional and state-of-the-art RL methods for traffic signal control in a synthetic traffic grid and three real-world traffic networks\footnote{\url{https://traffic-signal-control.github.io}} as follow:
    \begin{itemize}
    \itemsep=-1pt
        \item $D_{4\times4}$: The road network contains 16 intersections in a $4\times4$ grid. The traffic volume is randomly sampled from a Gaussian distribution with a mean of 500 vehicles/hour/lane.
        \item $D_{Jinan, Hangzhou}$: The road network of Jinan and Hangzhou contains 12 and 16 intersections in a $4\times3$ and $4\times4$ city network, respectively. The traffic flow is generated from surveillance camera data.
        \item $D_{Manhattan}$: The road network of Manhattan contains 48 intersections in a $16\times3$ city network. The number of vehicles generated is sampled from taxi trajectory data.
    \end{itemize}
    Specifically, we use the synthetic grid to benchmark our method under various flexible traffic patterns and real-world datasets to test the practicality of our method.

    We compare our AFMRL with the following conventional and leading RL methods:
    \begin{itemize}
    \itemsep=-1pt
        \item \textbf{SOTL} \cite{cools2013self} is controlled with additional demand responsive rules which compare the current phase with current traffic. It is a conventional method that utilizes current traffic.
        \item \textbf{MaxPressure} \cite{varaiya2013max} aims to control the intersection by balancing queue length between neighboring intersections by minimizing the ``pressure'' of the phases. It is the state-of-the-art conventional method for the network-level TSC.
        \item \textbf{CoLight} \cite{wei2019colight} uses graph attention network to facilitate information between neighbors.
        \item \textbf{PressLight} \cite{wei2019presslight} uses the max-pressure theory and designs the pressure as the reward of the agents, which has good performance in multi-intersection TSC.
        \item \textbf{MA2C} \cite{chu2019multi} uses multi-agent A2C to cooperatively control multi-intersections. And it includes information of neighborhood and spatial discount to stabilize the training process.
        \item \textbf{FMA2C} \cite{MWaamas20} is an extension of MA2C with feudal hierarchy for traffic signal control. It achieves better global cooperation through feudal hierarchy.
    \end{itemize}

    We implemented the experiments using the CityFlow simulator \cite{zhang2019cityflow} | a MARL environment for large-scale city traffic scenarios. For a fair comparison, we use the same experimental settings for specifying the Markov game model as in the FMA2C paper \cite{MWaamas20}. All the GNNs were implemented using the network structure in DiffPool \cite{ying2018hierarchical}.

	\subsection{Experimental Results}

	As commonly in the literature, we use the queue length and the average travel time as the metric to measure short-term and long-term traffic conditions respectively. The queue length is the number of queuing vehicles in the road network. The travel time is the time difference between the time when all cars enter the network and the one when they leave it. The average values were computed by several runs with different random seeds after convergence in training.

    The performance of all the methods in four traffic networks is summarized in Table \ref{tab:perf}, in terms of average travel time and average queue length. We can see that the traditional methods have relatively poor performance comparing to the RL approaches. As aforementioned, traditional methods that rely heavily on pre-defined traffic rules or assumptions do not work well with dynamic traffic patterns. Among the RL methods, FMA2C gained a slightly better performance due to the global cooperation brought by the feudal structure. All in all, our method achieved the best results especially for the large network (i.e., Manhattan).

    As shown in Figure \ref{Training-curves}(a-d), our method substantially outperformed all the compared RL methods, in the speed of convergence, the stability of learning, and the quality of policy. By comparing with FMA2C, we can find that our method with adaptive network partition converged faster and got better results in real-world traffic networks. This indicates the advantage of our method in situations where traffic patterns change frequently.

    Next, we will further investigate the reasons for this advantage through the ablation experiments.
	
	\subsection{Ablation Experiments}
	
	\begin{figure}[t]
		\centering
		\includegraphics[width=1.0 \hsize]{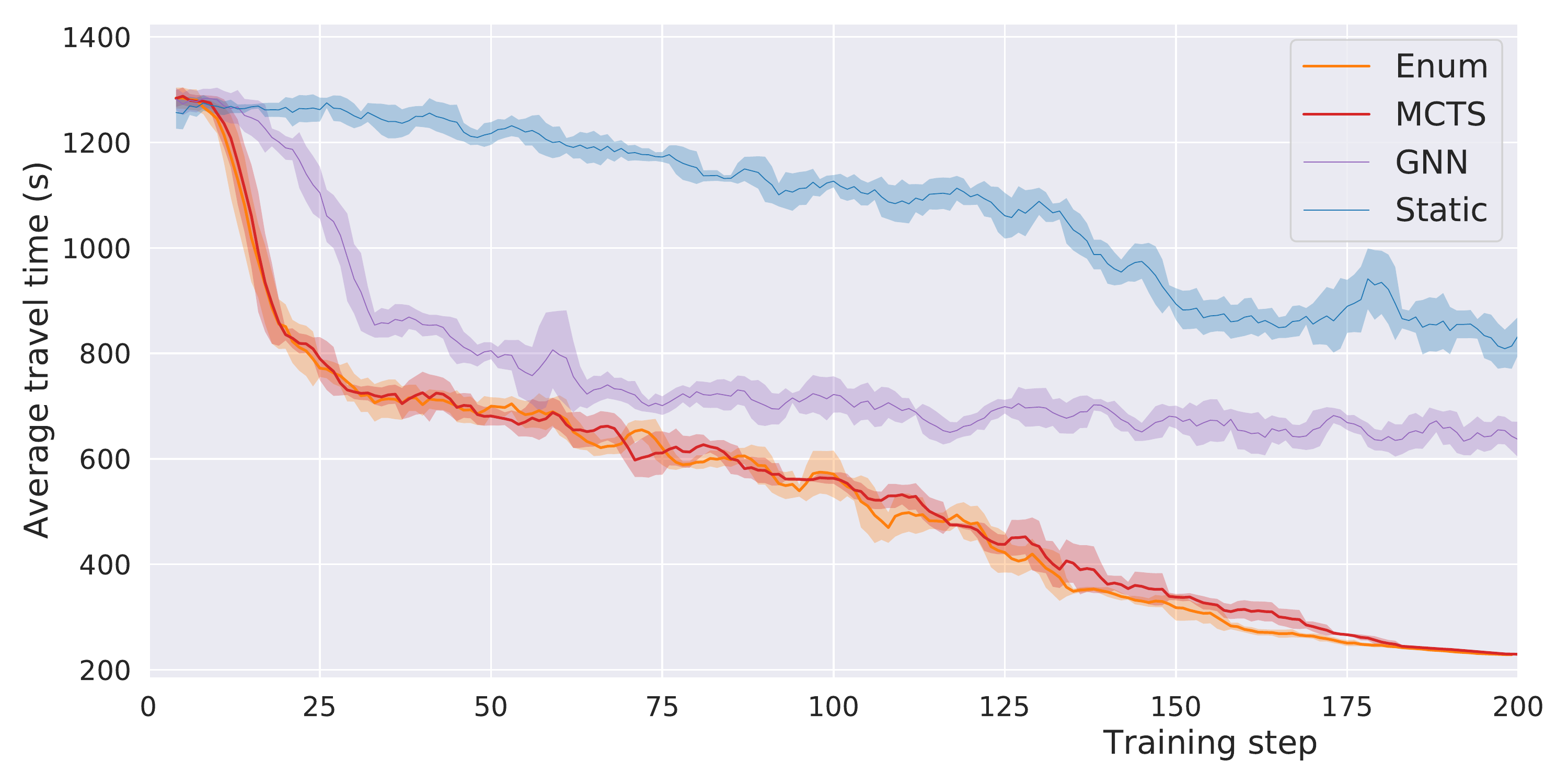}
		\caption{Ablation results for different partition methods.}
		\label{ablation}
	\end{figure}
	

    Here, we study the usefulness of different network partition techniques on the 4$\times$4 grid. Specifically, we tested the methods with static network partition (Static), GNN for network partition (GNN), and MCTS for network partition (MCTS). Exhaustive enumeration (Enum) for searching the best network partition is also included for reference though it is not scalable for large networks.
	
	\begin{figure}[t]
		\centering
		\subfigure[MCTS] {\includegraphics[width=0.32 \hsize]{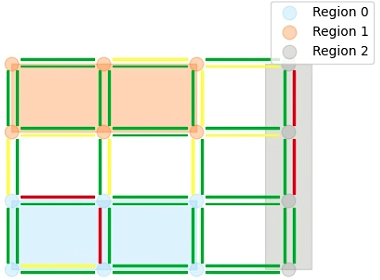}}
		\subfigure[Static] {\includegraphics[width=0.32 \columnwidth]{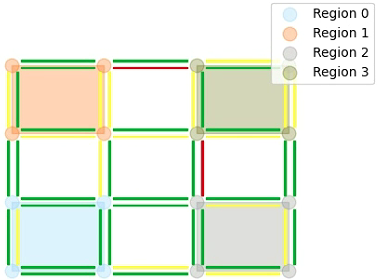}}
		\subfigure[GNN] {\includegraphics[width=0.32 \columnwidth]{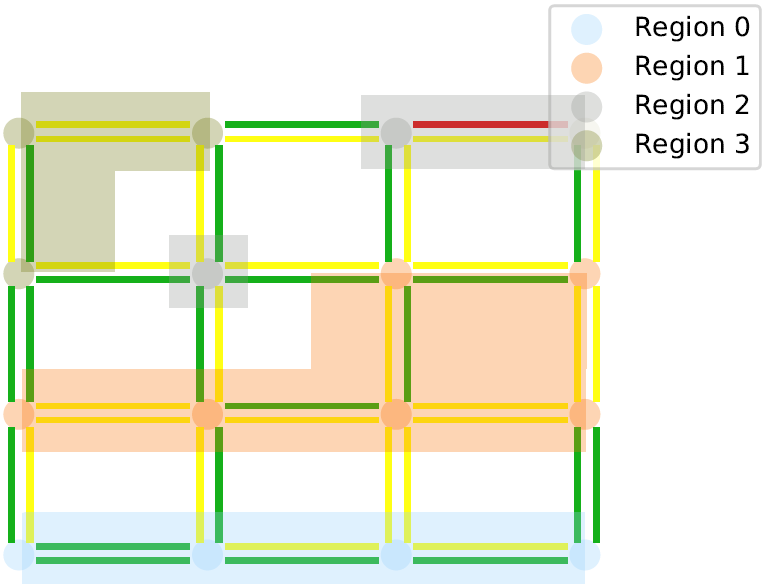}}
		\caption{Illustration of the network partition generated by different methods. Each intersection is represented by a vertex, and the connected edges represent the traffic flow density (green, yellow, and red indicate the degree of congestion). Different rectangles represent regions.}
		\label{ablation1}
	\end{figure}

	As shown in Figure \ref{ablation}, the static network partition has the worst performance when it does not match with the dynamic traffic flow. In more details, when the congested area spans two regions, the manager-level coordination is less effective. As shown in Figure\ref{ablation1}(b), the traffic density of the connection between the regions is high at this monment, e.g., the connection between the upper right region with its neighbors. In this case, the intra-regional connection is low, while the inter-regional connection is high. In contrast, GNN does adapt with the dynamic traffic flow. Therefore, its performance is better than the one with static partition. However, it is very hard to train and often fails to produce reasonable partition especially for large network. For example, as shown in Figure \ref{ablation1}(c), the intersections of the same region are not adjacent, which is usually a bad network partition for traffic signal control.

    For MCTS, as shown in Figure\ref{ablation1}(a), we can see that the partition is preferable, where the intra-regional connection is high, and the inter-regional connection is low. Moreover, this network partition will change dynamically along with the traffic flow\footnote{A sequence of changes is shown in the appendix.}. As a result, MCTS shows the best performance comparing with static and GNN for network partition. As we can see, when compared with enumeration baseline, MCTS got very close results to it. However, MCTS is much efficient and can scale to very large network. For instance, the time of the enumeration method for the 16$\times$3 network is 41.96s, while the time of MCTS is much less (i.e., 8.90s).
	
	\section{Related work}
	
	\subsubsection{Conventional traffic signal control.}
	
	In conventional control, most methods depend on hand-crafted rules \cite{koonce2008traffic, gershenson2004self}. 
	To solve difficult to manually design rules, max-pressure \cite{varaiya2013max} aims to balance the queue length between adjacent intersections by minimizing the "pressure" of the intersection. Overall, these approaches still rely on assumptions to simplify traffic conditions and do not guarantee optimal real-world results.
	
	\subsubsection{RL-based traffic signal control.}
	
	RL have been used for TSC so that the control strategy can be adaptively created based on the current traffic condition. 
	Some studies \cite{wiering2000multi, wei2018intellilight, zheng2019learning} relies on independent Q-learning (IQL). Obviously, they can't optimize the overall objective because they ignore the interaction between adjacent intersections in the road network.
	\citet{prashanth2010reinforcement} trains a central agent to control all intersections, and \citet{kuyer2008multiagent, van2016coordinated} jointly model the action with centralized optimization among agents. Due to the curse of dimensionality, these centralized methods are hard to apply on the large-scale road network.
	
	Other studies aim to encourage agents to learn cooperation with others in a decentralized way by adding information from neighbor \cite{aziz2018learning, wei2019colight, chu2019multi} or design a mechanism that induces cooperation \cite{wei2019presslight, xu2021hierarchically}.
	Due to the partial observation, these methods may be difficult to achieve global optimal control. Recently, FMA2C \cite{MWaamas20} combines MA2C with feudal hierarchy to improve global coordination among agents. We build our algorithm based on FMA2C and make adaptive network partition to adapt to more flexible and complex traffic.
	
	\section{Conclusion}
	
	In this paper, we propose a feudal MARL with adaptive network partition for TSC. First, we propose two approaches, based on GNN and MCTS respectively, to partition the traffic network into several regions fitting for dynamic traffic flow. 
	Then, we combine Qmix with GNN to extract the features of variant dimensions region and cooperative control the entire traffic network. Finally, we apply the feudal MARL to each region for global cooperation control.
	In the experiment, we benchmark in a synthetic and three real-world traffic networks. The results show better performance of our method over several state-of-the-art TSC methods and our advantage in the large-scale network. In the future, we plan to further improve the partition methods with prior knowledge and test it on real-world applications.
	
	\bibliography{main}
	
	\newpage
	\appendix
	
	In this Appendix, we will introduce the problem statement, model definition, network and hyperparameter settings of the experiments in details.

    \section{Problem Statement}
\begin{figure}[t]
    	\centering
    	\subfigure[An intersection and its 12 movements.] {\includegraphics[width=0.53 \hsize]{figure/intersection.png}}
    	\subfigure[Eight phases] {\includegraphics[width=0.45 \columnwidth]{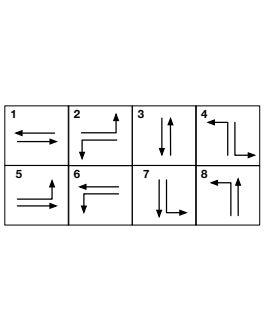}}
    	\caption{Example of the intersection and phases.}
    	\label{intersection}
    \end{figure}

    In this paper, we investigate traffic signal control in the scenario of multi-intersection. Similar to previous studies \cite{wei2019survey, chu2019multi}, we made the following definitions for an intersection, including road structure and traffic movement. An example is shown in Figure \ref{intersection}(a). These definitions can be easily extended to different types of intersections. Next, we will introduce the objective in the multi-intersection traffic signal control.
    
    \textbf{Road structure:} At the intersection, there are two kinds of traffic approaches: incoming and outgoing approaches. An incoming approach is that vehicles enter the intersection while an outgoing approach is that vehicles leave the intersection. There are four incoming approaches and four outgoing approaches, namely {\it East}, {\it South}, {\it West}, {\it North}, in the intersection. An approach consists of three lanes, namely {\it right-turn}, {\it straight} and {\it left-turn}.
    
    \textbf{Traffic movement and phase:} A traffic movement refers to vehicles moving from an incoming lane to an outgoing lane, such as the 12 movements shown in Figure \ref{intersection}(a). In daily traffic control, the green signal means that the movement is allowed to pass, and the red signal means the movement is prohibited. Generally, the combination of traffic movements for the green signal must be non-conflicting, and the combination is also called phase. As shown in Figure \ref{intersection}(a), phase \textbf{2} is set. There is a phase set containing eight phases for the traffic signal control as shown in Figure \ref{intersection}(b). 
    
    \textbf{Objective:} 
    There are many different metrics to measure traffic conditions. One of the most common metrics of traffic signal control is to minimize the {\em average travel time}. Average travel time is defined as the average time difference between the time when all vehicles enter the network and the time when they leave the network. However, because it can only be measured over a long time horizon, it is hard to optimize directly. In the experiments, we indirectly optimize average travel time by optimizing local travel time. The local travel time of an intersection is the time discrepancy between entering and leaving the local area of the intersection. And the local travel time of the traffic network is the sum of all intersections' local travel time.
    
    \section{Model Definition}
    Here, we introduce the observation, action, and reward definition for the feudal structure, i.e., manager-worker hierarchy. There are many different definitions in the literature \cite{wei2019survey}. In this paper, we refer to the definitions of \cite{MWaamas20},  with the only difference of adjustment in the manager settings.
    
    \textbf{Observation:} For worker $i$, the observed traffic information is some quantitative  descriptions of the intersection i, i.e., queue length, waiting time and delay. In this paper, we choose queue length of all lane in the intersection as the observation, $o^W_{t, i} = \langle q_1, \cdots, q_{l_n}\rangle$, where $l_n$ is the number of lanes in intersection $i$. For manager $k$, the observation is the traffic flow in the region $k$, i.e., $o^M_{t, k} = \langle ..., o^W_{t, i}, ... \rangle$, where $i \in Region_k$. And the region's observation will be input into a GNN to extract the its feature emmbddings.
    
    \textbf{Action:} For each worker, it decides which phase to be selected from a phase set. In other words, the action is the index of a phase. For each manager, it sets a sub-goal for its region. And we consider the sub-goal as a possible traffic flow, which is a combination of $[N, E, W, S]$ traffic flows, e.g, {\it north-south} and {\it east-west} traffic flows.
    
    \textbf{Reward:} For worker $i$, the reward received from environment is the queue length of all lanes in the intersection $i$, which can be represented as $r_{i} = - \sum_{l=1}^{l_n} q_l$. For manager $k$, we define a reward of long-time horizon. Note that we employ Qmix \cite{rashid2018qmix} to estimate joint action-values as a complex non-linear combination of per-manager values that promote the same optimization objective of managers. We consider the local travel time, which is the sum of local travel time of all intersections of the traffic network. The local travel time of an intersection is the time discrepancy between entering and leaving the local area of the intersection. Therefore, the manager can indirectly optimize the average travel time by optimizing the local travel time.
    
    \section{Network And Hyperparameters}
    In this section, we first describe more details of the GNNs and Qmix. Note that we have used three GNNs in the paper. Then, we introduce their network structure and training process in details.
    
    \begin{figure}[t]
    	\centering
    	\includegraphics[width=1.0 \hsize]{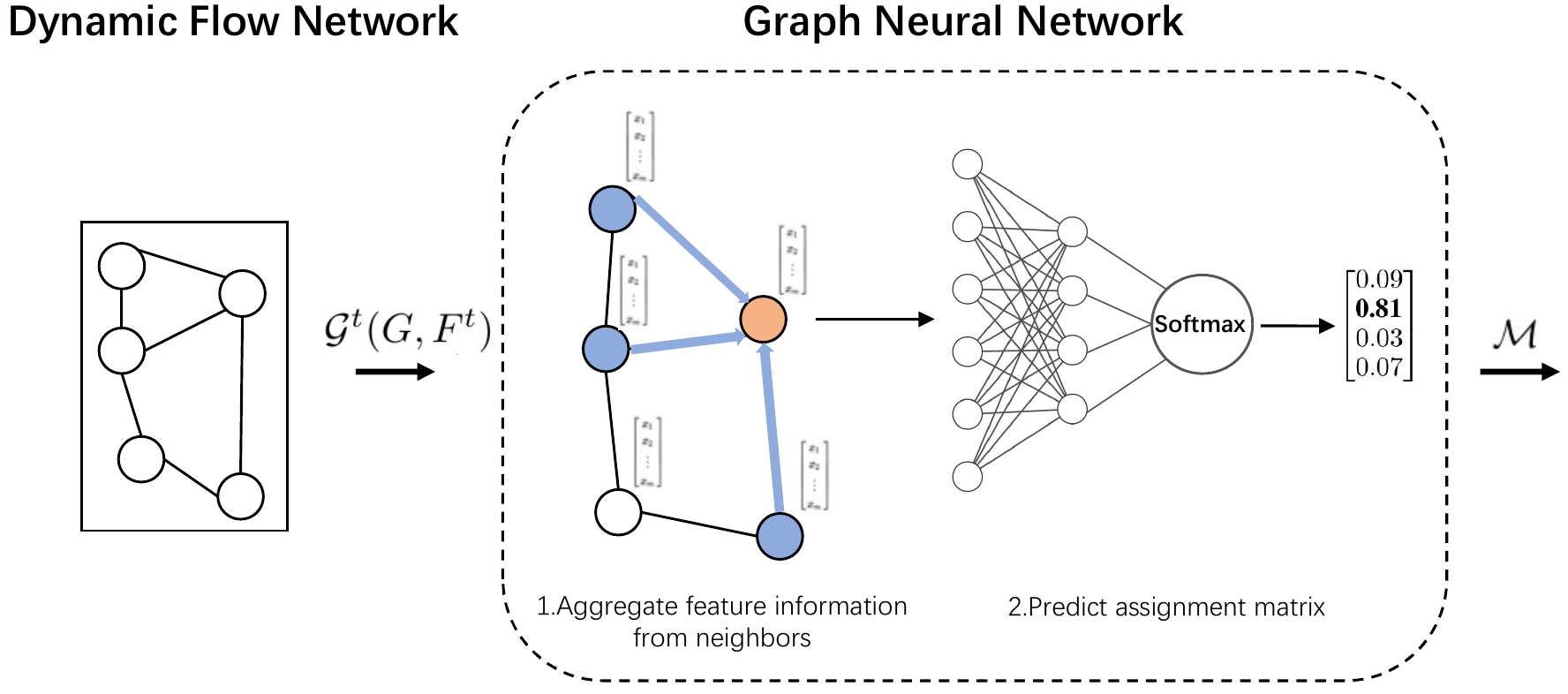}
    	\caption{Illustration of the GNN for generating network partition $\mathcal{P}$ (The colored arrow line represents the process that aggregate feature information from neighbors).}
    	\label{network1}
    \end{figure}
    
    \subsubsection{GNN for network partition.} Firstly, we introduce our method of applying the GNN to generate the network partition $\mathcal{P}$ given the dynamic flow network. As shown in Figure \ref{network1}, the GNN takes the $\mathcal{G}^t(G, F^t)$ as input, which contains the observation of intersections $\mathcal{O}$ and the real-time flow $F^t$. The first step is to generate embedding of nodes and aggregate feature information from the neighbors (blue) to the target node (red). Then we input the embedding of the target node into an MLP with $softmax$ to get the assignment vector, which can be regarded as a multi-classifier. The output dimension of the GNN corresponds to the number of regions. Note that the GNN used by us refers to $GNN_{pool}$ mentioned in DiffPool \cite{ying2018hierarchical}.
    
    Our objective of training the GNN is to generate a network partition with a good collective payoff that the agents can gain by forming the network partition in TSC. As shown in Figure 2 in the main body, we train the GNN in an end-to-end fashion using the gradient signals, which are backpropagated by the loss function of Qmix, i.e., Equation (9). In practice, it is generally difficult to train the GNN using only gradient signals from the RL. Hence, we use an auxiliary link loss to encode nearby nodes that should be pooled together as suggested by \citet{ying2018hierarchical}, i.e., minimizing the loss $\mathcal{L}_{aux} = \|A, \mathcal{M} \mathcal{M}^T\|$, where $\|\|$ is the Frobenius norm and $A$ is the adjacency matrix of the traffic network. In summary, we train the GNN in an end-to-end fashion with the Loss function $\mathcal{L} = \mathcal{L}_{aux} + \mathcal{L}_{Qmix}$.
    
    \begin{figure}[t]
    	\centering
    	\includegraphics[width=1.0 \hsize]{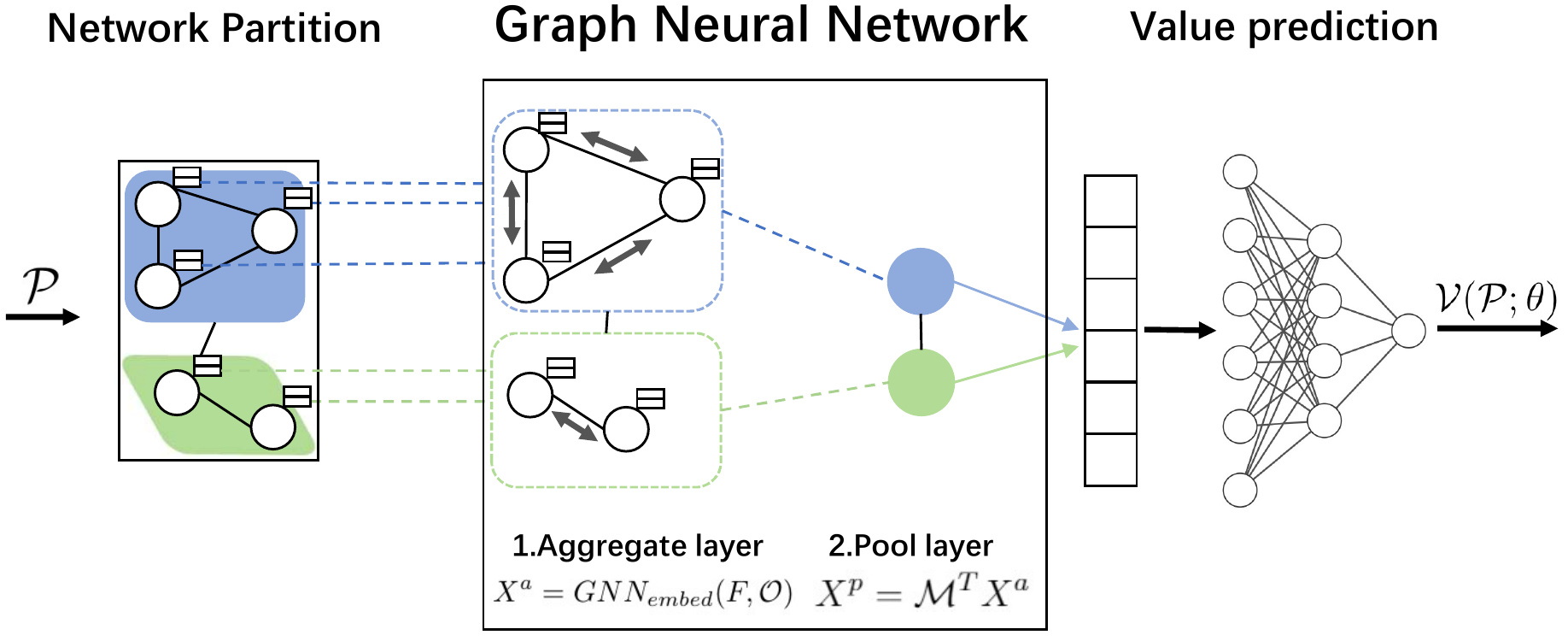}
    	\caption{Illustration of predicting network partition value $\mathcal{V}(\mathcal{P};\theta)$ by GNN ( The colored rectangle represents a region, and the black line represents the traffic flow between them).}
    	\label{network2}
    \end{figure}
    
    \subsubsection{GNN for partition evaluation in MCTS.} Secondly, we introduce another GNN for predicting the value function $\mathcal{V}(\mathcal{P})$ of $\mathcal{P}$ given any network partition $\mathcal{P}$. Note that this is a separate GNN that is structurally different from the aforementioned GNN for network partition. As shown in Figure \ref{network2}, the network takes the observation of agents $\mathcal{O}$, the real-time flow $F^t$ (the relation between a pair of agents) and the assignment matrix $\mathcal{M}$ as inputs, and outputs the embeddings for the regions. As mentioned in the main body, we sequentially generate the embeddings $X^a$ of agents, coarsened adjacency matrix $\mathcal{F}$ and the embeddings ${\hat{X}^p}$ for regions using Equation (3-6). Finally, the embeddings ${\hat{X}^p}$ feed in the neural network to output the value $\mathcal{V}(\mathcal{P};\theta)$. 
    
    To train the GNN, we expect to fit $\mathcal{V}(\mathcal{P}; \theta)$ as the collective payoff that the agents can gain by forming the network partition $\mathcal{P}$ in TSC. Therefore, we use the rewards $r$, which is received from the environment by interacting with the environment during $T$ steps, as the training labels. Since the high-level reward $r$ represents the local travel time of the traffic network, it can be used to represent how much collective payoff that the agents can gain by forming the network partition $\mathcal{P}$. To put together, the network is trained in an end-to-end fashion by minimizing $\mathcal{L}$ = MSE($\mathcal{V}(\mathcal{P};\theta), r$).
    
    \begin{figure}[t]
    	\centering
    	\includegraphics[width=1.0 \hsize]{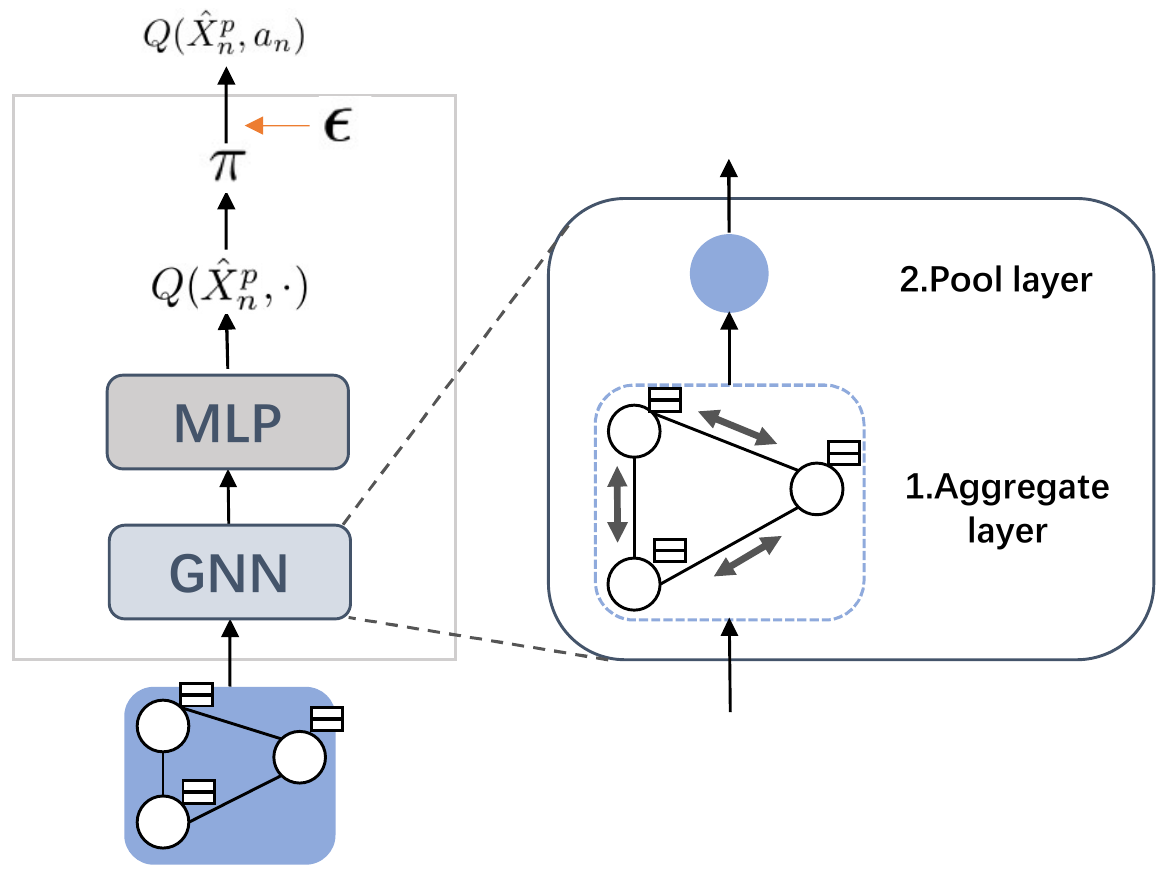}
    	\caption{Illustration of the GNN for computing inputs of independent Q-value network in Qmix.}
    	\label{qmix}
    \end{figure}
    
    \begin{figure*}[t]
    	\centering
    	\includegraphics[width=2.0\columnwidth]{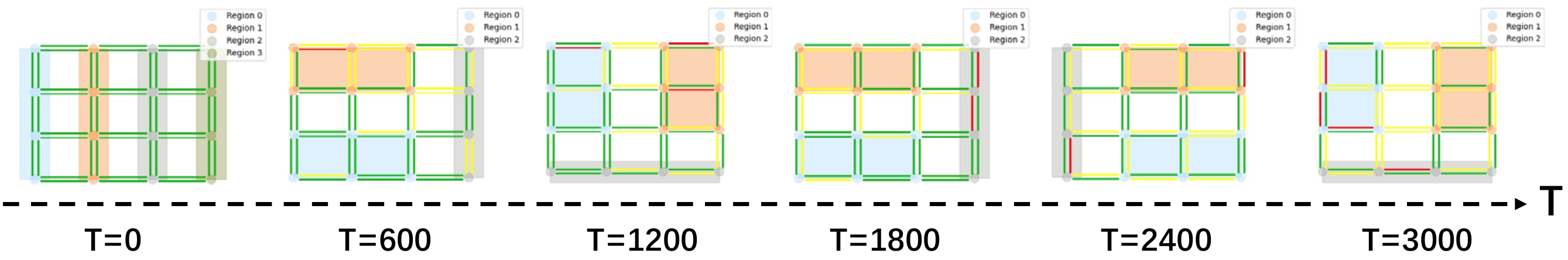}
    	\caption{Illustration of the sequence of network partitions changed dynamically along with the traffic flow in an episode. Each intersection is represented by a vertex, and the connected edges represent the traffic flow density (green, yellow, and red indicate the degree of congestion). Different rectangles represent different regions.}
    	\label{map}
    \end{figure*}
    
    \begin{figure*}[t]
    		\centering
    		\subfigure[Synthetic 4x4] {\includegraphics[width=0.9\columnwidth]{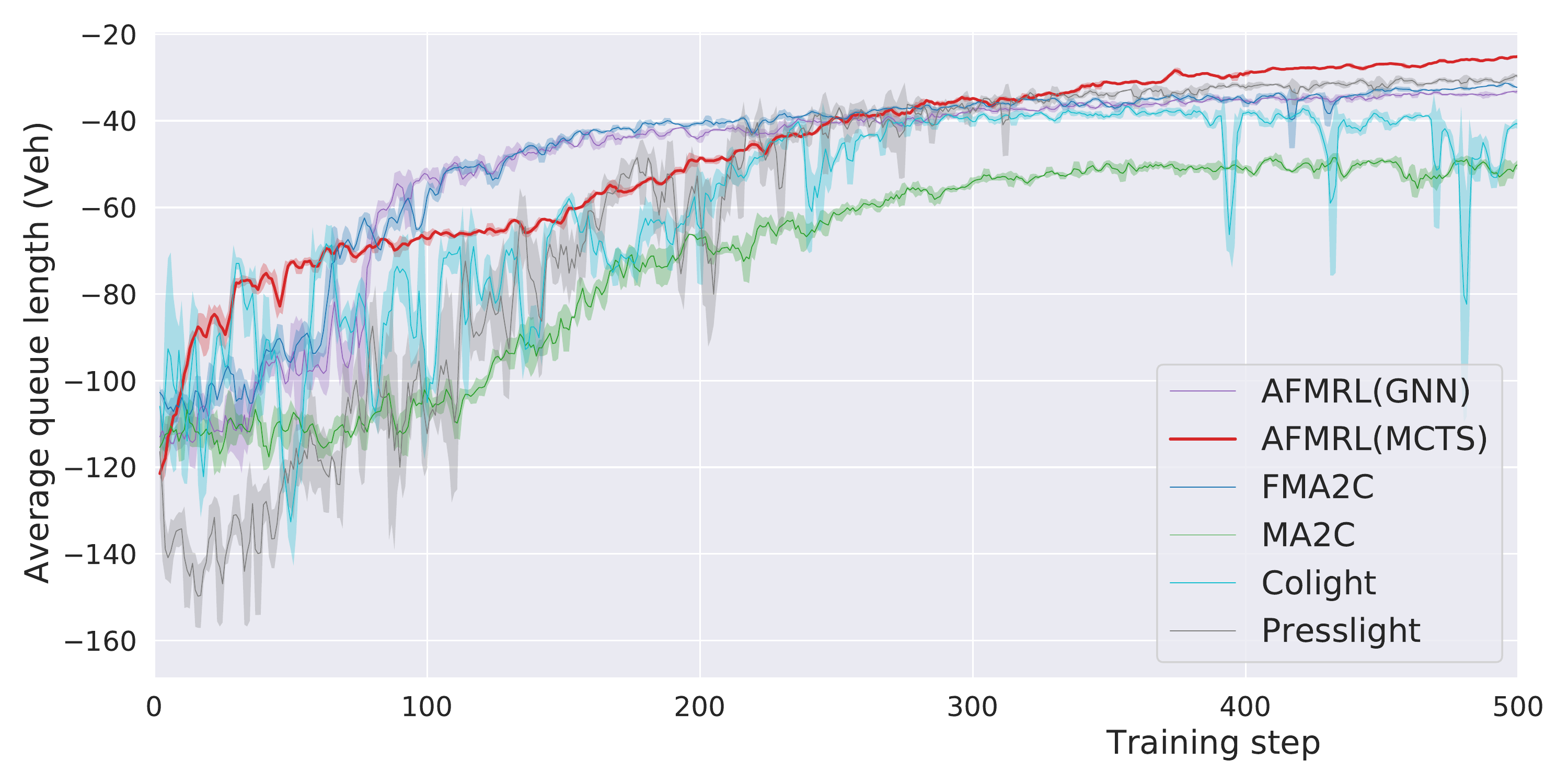}}
    		\subfigure[Jinan] 
    		{\includegraphics[width=0.9\columnwidth]{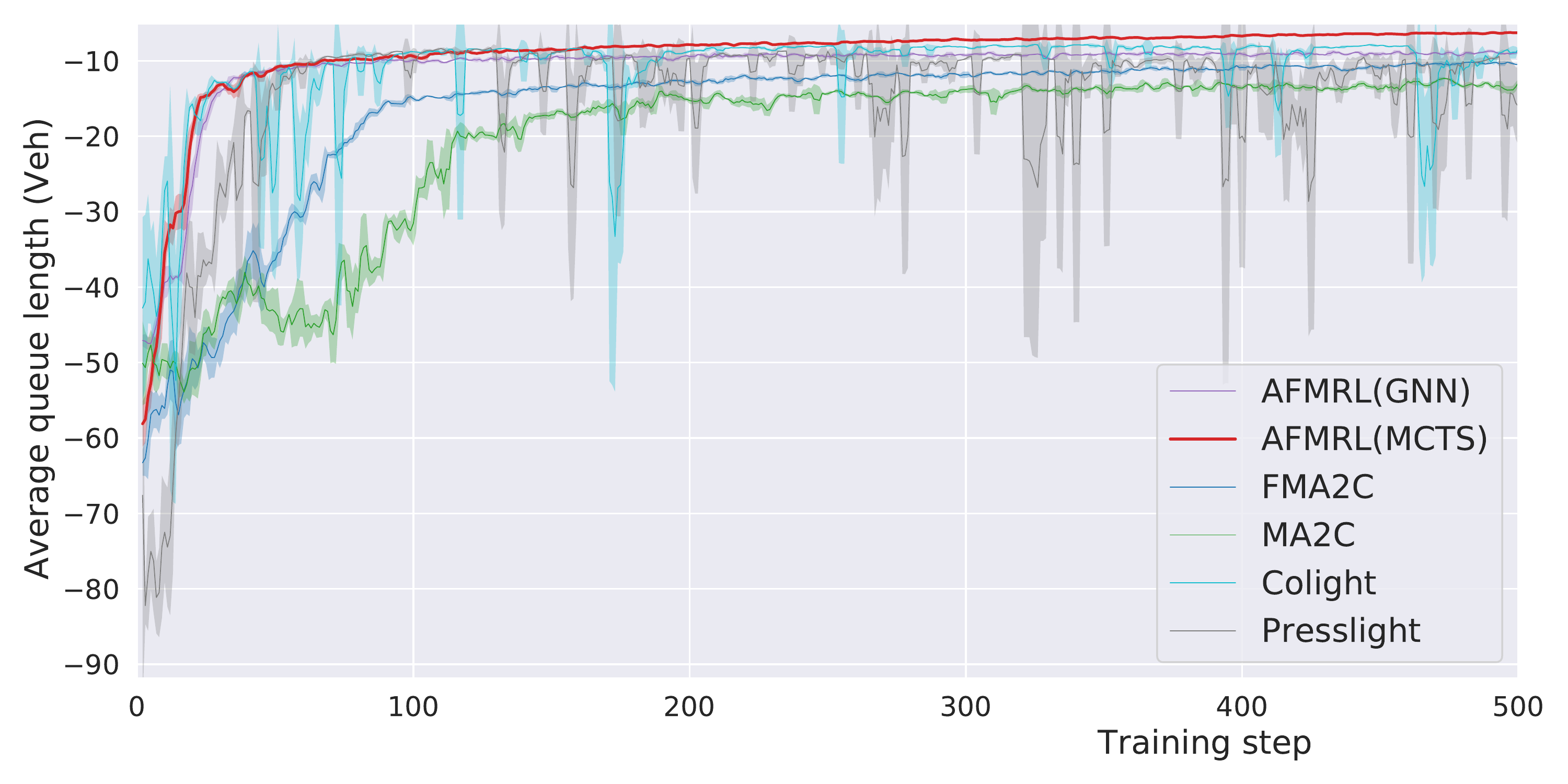}}
    		\subfigure[Hangzhou] {\includegraphics[width=0.9\columnwidth]{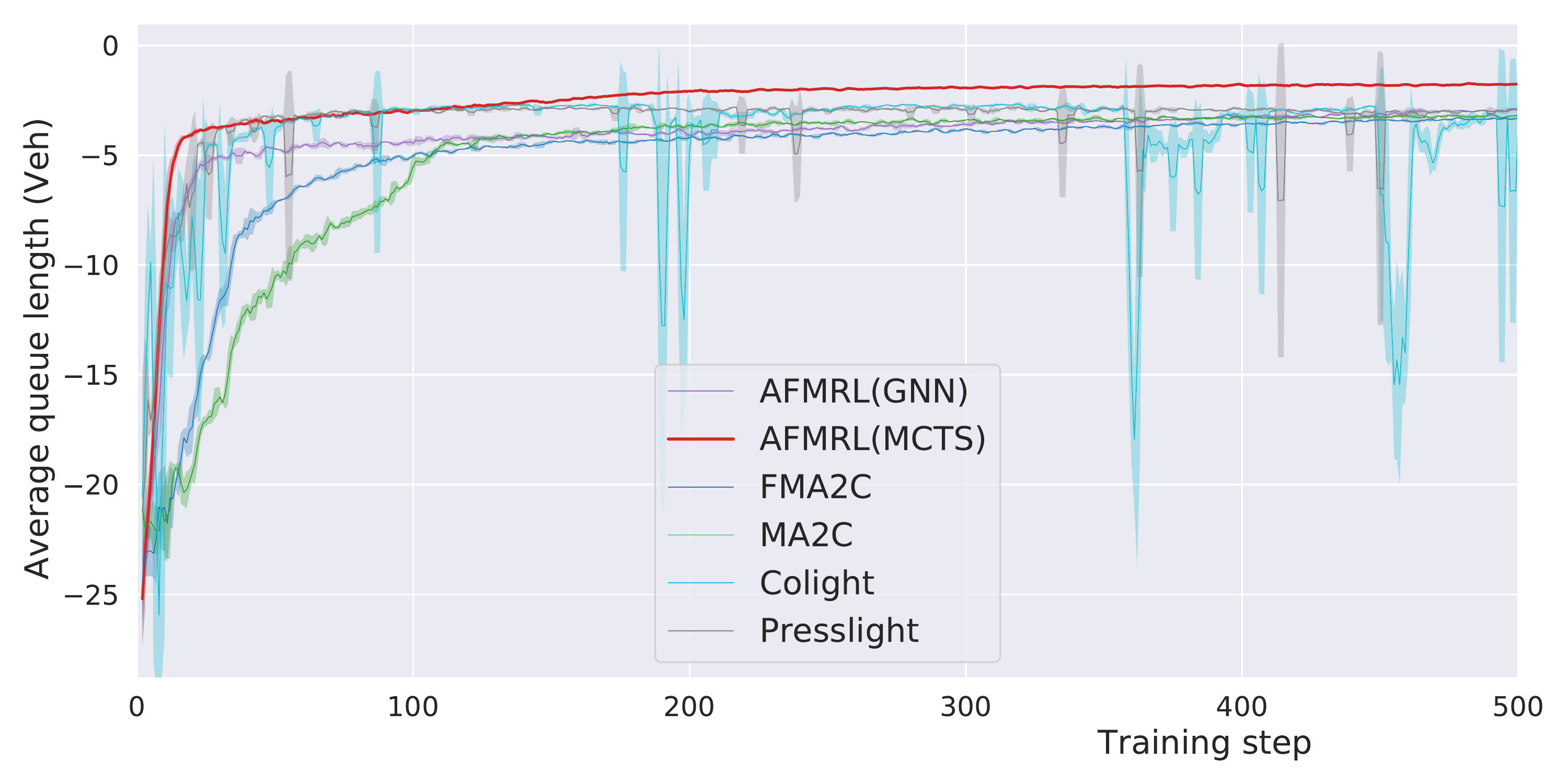}}
    		\subfigure[Manhattan] {\includegraphics[width=0.9\columnwidth]{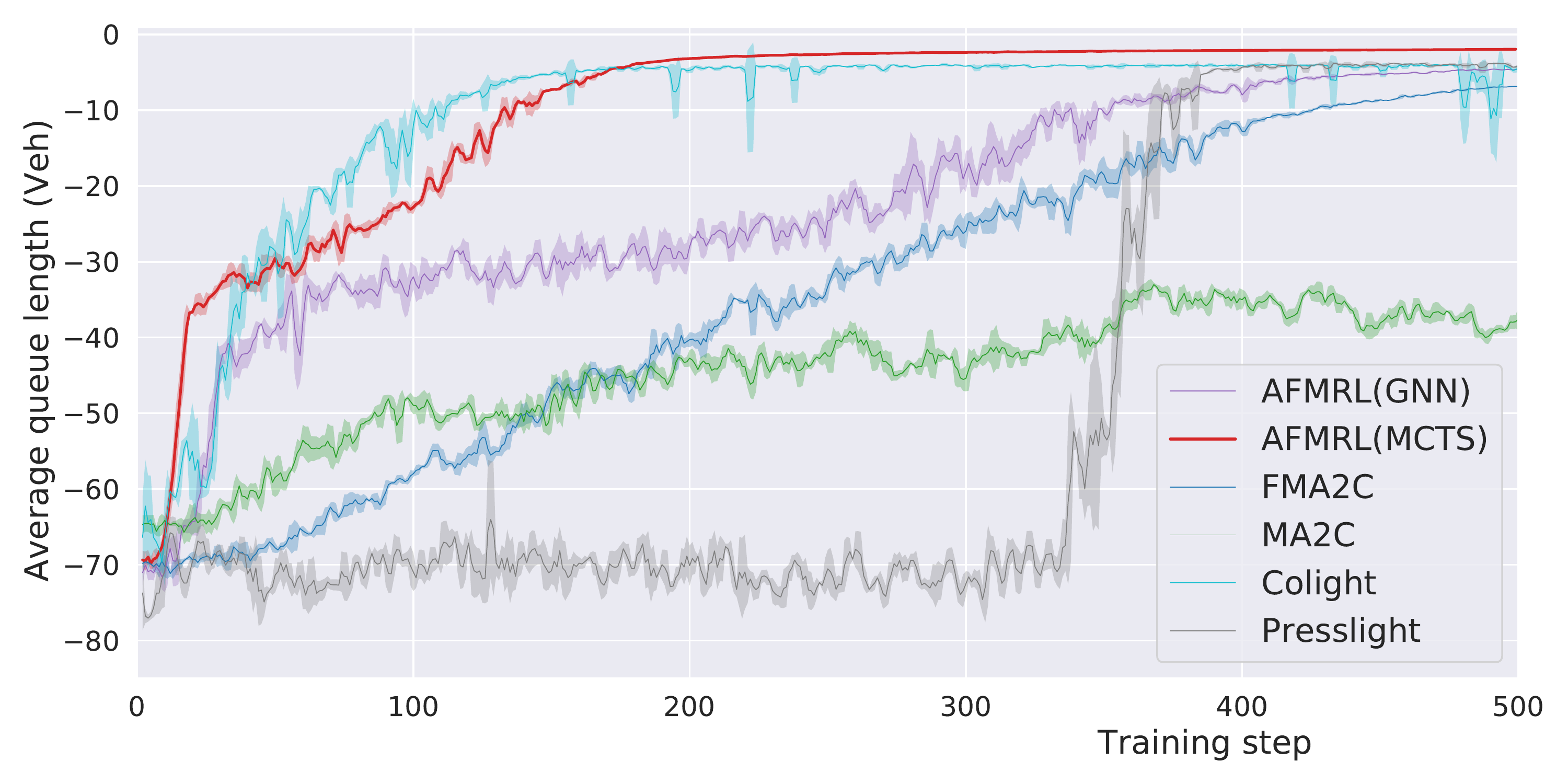}}
    		\caption{The train curves of average queue length (the solid line and shade are the mean and standard deviation respectively).}
    		\label{Training-curves-q}
    \end{figure*}
    
    \subsubsection{GNN for embedding partitions in Qmix.} Thirdly, we introduce the detail of the GNN used by the independent Q-value network in Qmix. In this GNN, each manager controls a region, so the input is the observation of the region, which includes the observation of workers in the region. Notice that, for the different situations of the traffic network, there will be different network partitions, i.e., the number of regions and the number of agents in the region. Due to the dynamic dimensions of input, the region managed by the manager may be different. Therefore, we use a GNN to extract features of each region. By doing so, we can use a unified Q-value network to process dynamic dimensions of input and output the individual Q-value function $<Q^1, ..., Q^m>$. For the mixing network, we still adopt the method introduced in Qmix \cite{rashid2018qmix}. The hyper-parameter of the mixing network takes the environment state $s$ as input and generates the non-negative weights and bias of the layer in the mixing network. And the mixing network takes the individual Q-value function $<Q^1, ..., Q^m>$ to generate the total Q-value function $Q^{tot}$. The GNN is trained in an end-to-end manner using the Qmix loss function, i.e., Equation (9).
    
    Above, we have summarized three different GNNs used in our method and introduced their structure and training methods. In our experiments, we refer to the structure of the three GNNs in DiffPool \cite{ying2018hierarchical}.
    
    \subsubsection{Hyperparameters for the experiments.}Table \ref{Hyperparameters} summarizes the hyperparameters of our AFMRL, FMA2C\cite{MWaamas20}, MA2C\cite{chu2019multi}, Colight\cite{wei2019colight} and PressLight\cite{wei2019presslight}. For the baselines, we use their open-source implementations and their default parameter settings for fair comparison.
    
    \begin{table}[htp]
    \centering \small
    \caption{Hyperparameters of ours and other methods.}
    \begin{tabular}{p{70pt} p{120pt} c}
    \toprule
    \textbf{Algorithms} & \textbf{Hyperparameters} & \textbf{Value} \\ \hline
    \multirow{12}{*}{AFMRL}
    & discount factor $\gamma$ & 0.99 \\
    ~ & entropy regularization $\beta$ & 0.01 \\
    ~ & decay & Linear \\
    ~ & $T$ & 30 \\
    ~ & optimizer & RMSProp \\
    ~ & learning rate $\eta$ & 5e-4 \\
    ~ & \# MLP layers of Worker & 2 \\
    ~ & \# MLP units of Worker & (64, 32) \\
    ~ & \# LSTM units of Worker & (64) \\
    ~ & \# MLP layers of Qmix & 2 \\
    ~ & \# MLP units of Qmix& (64, 64) \\
    ~ & MLP activation & ReLU \\
    ~ & initializer & Random \\
    \midrule
    \multirow{12}{*}{FMA2C}
    & discount factor $\gamma$ & 0.99 \\
    ~ & entropy regularization $\beta$ & 0.01 \\
    ~ & decay & Linear \\
    ~ & $T$ & 30 \\
    ~ & optimizer & RMSProp \\
    ~ & learning rate $\eta$ & 5e-4 \\
    ~ & \# MLP layers of Worker & 2 \\
    ~ & \# MLP units of Worker & (64, 32) \\
    ~ & \# LSTM units of Worker & (64) \\
    ~ & \# MLP layers of Manager & 2 \\
    ~ & \# MLP units of Manager & (64, 32) \\
    ~ & \# LSTM units of Worker & (64) \\
    ~ & MLP activation & ReLU \\
    ~ & initializer & Random \\
    \midrule
    \multirow{10}{*}{MA2C} & discount factor $\gamma$ & 0.99 \\
    ~ & entropy regularization $\beta$ & 0.01 \\
    ~ & decay & Linear \\
    ~ & optimizer & RMSProp \\
    ~ & learning rate $\eta$ & 5e-4 \\
    ~ & \# MLP layers & 2 \\
    ~ & \# MLP units & (64, 32) \\
    ~ & \# LSTM units & (64) \\
    ~ & MLP activation & ReLU \\
    ~ & initializer & Random \\
    \midrule
    \multirow{10}{*}{Colight, PressLight}
    & discount factor $\gamma$ & 0.8 \\
    ~ & batch $\beta$ & 0.01 \\
    ~ & buffer capacity & 1000 \\
    ~ & sample size & 30 \\
    ~ & decay & 0.95 \\
    ~ & optimizer & RMSprop \\
    ~ & learning rate $\eta$ & 1e-3 \\
    ~ & \# MLP layers of Worker & 4 \\
    ~ & \# MLP units of Worker & (32, 32) \\
    ~ & MLP activation & ReLU \\
    ~ & initializer & Random \\
    \bottomrule
    \end{tabular}
    \label{Hyperparameters}
    \end{table}
    
    \section{Additional Experimental Results}
    
    This section shows some additional experimental results. In the experiments, we use the queue length and the average travel time as the metric to measure traffic conditions, for short-term and long-term traffic conditions respectively. In the main body of the paper, we showed the train curves of the average travel time. Here, we report our results of the average queue length of the intersections in the tested networks. As shown in Figures \ref{Training-curves-q}(a-d), our method substantially outperformed all the compared methods, both in the speed of convergence, the stability of learning, and the quality of policy. In summary, our method can optimize either short-term or long-term goals in traffic signal control very well.
    
    As shown in Figure \ref{map}, we show a sequence of network partitions, which dynamically fit the real-time traffic flow. We can see that: the traffic flow changed over time, and the traffic density of the connection between a pair intersections is various at different moments. Thanks to our adaptive network partition, as expected, the inter-regional connection is relatively low, while the intra-regional connection is relatively high. Intuitively, this is a reasonable network partition and beneficial for learning good policies for the agents. This demonstrates that our method is effective for multi-intersections traffic signal control.
	
\end{document}